\documentclass[11pt,a4paper]{article} 
\usepackage{amsmath}
\usepackage{graphicx}
\usepackage{hyperref}
\usepackage{bm}
\usepackage{authblk}
\usepackage[numbers,sort&compress]{natbib}
\hypersetup{bookmarks,colorlinks,linkcolor=[rgb]{0.2,0.2,0.2},citecolor=magenta}
\reversemarginpar
\newcommand\TT{\rule{0pt}{2.5ex}}        
\newcommand\BB{\rule[-1.0ex]{0pt}{0pt}}  

\newcommand{\be}{\begin{equation}}
\newcommand{\en}{\end{equation}}
\newcommand{\bea}{\begin{eqnarray}}
\newcommand{\ena}{\end{eqnarray}}
\newcommand{\bp}{\begin{pmatrix}}
\newcommand{\ep}{\end{pmatrix}}
\newcommand{\lbl}[1]{\label{eq:#1}}

\newcommand{\lblfig}[1]{\label{fig:#1}}

\newcommand{\rf}[1]{(\ref{eq:#1})}

\newcommand{\fig}[1]{\ref{fig:#1}}
\newcommand{\sect}[1]{\ref{sec:#1}}
\newcommand{\braque}[1]{{\langle #1 \rangle}}
\newcommand{\ket}[1]{{\vert #1 \rangle}}
\newcommand{\unun}{{{1\over2}{1\over2}}}
\newcommand{\unmun}{{{1\over2}{-1\over2}}}

\newcommand{\bc}{\begin{center}}
\newcommand{\ec}{\end{center}}
\newcommand{\bt}{\begin{tabular}}
\newcommand{\et}{\end{tabular}}
\newcommand{\ba}{\begin{array}}
\newcommand{\ea}{\end{array}}
\newcommand{\gapprox}{%
\mathrel{%
\setbox0=\hbox{$>$}\raise0.6ex\copy0\kern-\wd0\lower0.65ex\hbox{$\sim$}}}
\newcommand{\lapprox}{%
\mathrel{%
\setbox0=\hbox{$<$}\raise0.6ex\copy0\kern-\wd0\lower0.65ex\hbox{$\sim$}}}
\newcommand{\inleft}{%
\mathrel{%
\setbox0=\hbox{$<$}\copy0\kern-0.5\wd0\lower1.1\ht0\hbox{$\scriptstyle{in}$}}}
\newcommand{\inright}{%
\mathrel{%
\setbox0=\hbox{$>$}\copy0\kern-0.5\wd0\lower1.1\ht0\hbox{$\scriptstyle{in}$}}}
\newcommand{\outleft}{%
\mathrel{%
\setbox0=\hbox{$<$}\copy0\kern-0.5\wd0\lower1.1\ht0\hbox{$\scriptstyle{out}$}}}
\newcommand{\outright}{%
\mathrel{%
\setbox0=\hbox{$>$}\copy0\kern-0.5\wd0\lower1.1\ht0\hbox{$\scriptstyle{out}$}}}
\newcommand{\MeV}{\hbox{MeV}}
\newcommand{\im}{{\rm Im\,}}
\newcommand{\re}{{\rm Re\,}}
\newcommand{\disc}{{\rm disc\,}}
\newcommand{\Kbar}{\bar{K}}
\newcommand{\mpi }{m_\pi}
\newcommand{\fpi }{F_\pi}
\newcommand{\meta }{m_\eta}
\newcommand{\mpid}{m_\pi^2}

\newcommand{\fpid}{F_\pi^2}
\newcommand{\mpiq}{m_\pi^4}
\newcommand{\mkd}{m_K^2}

\newcommand{\Kp}{{K^+}} 
\newcommand{\Km}{{K^-}}
\newcommand{\Kz}{{K^0}} 
\newcommand{\Kzb}{{\bar{K}^0}}
\newcommand{\piz}{{\pi^0}} 
\newcommand{\pip}{{\pi^+}}
\newcommand{\pim}{{\pi^-}}
\newcommand{\metad}{m_\eta^2}
\newcommand{\msv}{-s_V}
\newcommand{\psv}{s_V}
\newcommand{\mvd}{m_V^2}
\newcommand{\mv}{m_V}
\newcommand{\sbarv}{{{s}_V}}
\newcommand{\sbark}{{s_{K^*}}}

\title {\bf The $\pi\eta$ interaction and $a_0$ resonances  in photon-photon
  scattering}

\author[a,b]{Junxu Lu}

\affil[a]{\small P\^ole Th\'eorie, IJCLab (CNRS/IN2P3, UMR9012), Universit\'e
  Paris-Saclay, 91406 Orsay, France}

\affil[b]{\small School of Physics, Beihang
  University, Beijing 102206, China}

\author[a]{B. Moussallam}

\begin{document}
\maketitle

\begin{abstract}
We revisit the information on the two lightest $a_0$ resonances and 
$S$-wave $\pi\eta$ scattering that can be extracted from
photon-photon scattering experiments. 
For this purpose we construct a model for the $S$-wave photon-photon
amplitudes which satisfies analyticity properties, two-channel unitarity and
obeys the soft photon as well as the soft pion constraints. The underlying
I=1 hadronic $T$-matrix involves six phenomenological parameters and is able
to account for two resonances below 1.5 GeV.
We perform a combined fit of the $\gamma\gamma\to \pi\eta$ and
$\gamma\gamma\to K_SK_S$ high statistics experimental data from the
Belle collaboration.  
Minimisation of the $\chi^2$ is found to have two distinct solutions with
approximately equal $\chi^2$. One of these exhibits a light and narrow excited
$a_0$ resonance analogous to the one found in the Belle analysis. This however
requires a peculiar coincidence between the $J=0$ and $J=2$ resonance effects
which is likely to be unphysical.  In both solutions the $a_0(980)$ resonance
appears as a pole on the second Riemann sheet. The location of this 
pole in the physical solution is determined to be
$m-i\Gamma/2=1000.7^{+12.9}_{-0.7} -i\,36.6^{+12.7}_{-2.6}$ MeV.
The solutions are also compared to experimental data in the
kinematical region of the decay $\eta\to\pi^0\gamma\gamma$. In this
region an isospin violating contribution associated with
$\pip\pim$ rescattering must be added for which we provide a
dispersive evaluation.

\end{abstract}
\newpage

\tableofcontents

\section{Introduction:}
The $a_0(980)$ and $f_0(980)$ scalar mesons were the first observed members
of a family of exotic resonances in QCD which are located very close to an
inelastic two-particle (or quasi two-particle) threshold (see the
review~\cite{Klempt:2007cp}).  The $a_0(980)$ resonance was discovered a long
time ago and seen in both the $K\Kbar$ and the $\pi\eta$
channels~\cite{Astier:1967zz,Ammar:1969vy} but its properties are still
imprecisely known. This is partly because $\pi\eta$ production experiments
using an $\eta$ beam, analogous to those which have allowed to determine the
$\pi\pi$ or $\pi K$ phase shifts, are not feasible. The $a_0(980)$ properties
have to be determined solely from final-state rescattering effects. As shown
by Flatt\'e~\cite{Flatte:1976xu}, who proposed a simple replacement for the
Breit-Wigner resonance formula accounting for the $\pi\eta-K\Kbar$
coupled-channel dynamics, ambiguous results for the $a_0$ width can be
obtained. In the PDG~\cite{Tanabashi:2018oca}, indeed, its width is simply
quoted as lying in a range from 50 to 100 MeV. Beyond the value of the width,
one would like to determine the positions of the poles on the unphysical
Riemann sheets. For resonance states close to an inelastic threshold, there
are three sheets (II, III, IV in the case of the $a_0$) which are physically
relevant (i.e. a pole in one of these can be close to the physical region). It
is also clear that a better determination of the resonance properties is
closely tied to a better knowledge of the physical scattering $T$-matrix.

A theoretically motivated treatment of final-state interactions becomes
prohibitively difficult for multiparticle final states. In this regard,
photon-photon to meson-meson scattering amplitudes are very favourable
processes\footnote{A renewed interest in such processes, both
  theoretical and experimental, is motivated, more generally, by the
  problem of evaluating the so-called hadronic light-by-light
  contribution to the muon $g-2$ (see
  e.g.~\cite{Jegerlehner:2009ry}).}.
They are free of initial-state interactions and satisfy dispersion
relations which can be constrained, in the case of $\pi\pi$ or $\pi\eta$ by
both soft photon~\cite{Low:1958sn,Abarbanel:1967wk} and soft
pion~\cite{Adler:1965ga} low-energy theorems. As was illustrated in the
seminal papers~\cite{Morgan:1987wi,Morgan:1991zx} a predictive representation
can be implemented at the level of the partial-waves with a simple modelling
of the left-hand cut.
From an experimental point of view $\gamma\gamma\to \eta\pi^0$
cross-sections were first measured by the Crystal Ball
collaboration~\cite{Antreasyan:1985wx}.  Measurements with much higher
statistics were recently performed by the Belle
collaboration~\cite{Uehara:2009cf}. There are also experimental results on the
$\eta\to \pi^0\gamma\gamma$ decay
amplitude~\cite{Prakhov:2008zz,Nefkens:2014zlt}.  Recent experimental data
exist also for $\gamma\gamma\to \Kz\Kzb$~\cite{Uehara:2013mbo} and
$\gamma\gamma\to \Kp\Km$~\cite{Abe:2003vn} which will be considered in our
study.

There are two puzzling aspects in the data analysis performed by the Belle
collaboration which we wish to reconsider: a) they find that the $a_0(980)$
peak seems to be best described by an ordinary (i.e. essentially elastic)
Breit-Wigner function and b) they find an excited $a_0$, which could
correspond to the $a_0(1450)$, but has a width, $\Gamma=65^{+2.1}_{-5.4}$ MeV,
much smaller than the PDG average as well as a significantly smaller
mass. These data have been re-analysed recently~\cite{Danilkin:2017lyn} based
on a specific meson-meson $T$-matrix model~\cite{Danilkin:2011fz} applied to
the $\pi\eta$ $S$-wave, from which the corresponding $\gamma\gamma$ amplitude
is deduced from a Muskhelishvili-Omn\`es (MO)
construction~\cite{Muskhelishvili,Omnes:1958hv}. Including also the
$a_2(1320)$ resonance but no $a_0(1450)$ resonance a good description of the
data up to 1.1 GeV and a qualitative description up to 1.4 GeV has
been achieved. The $a_0(980)$ pole, in this model, lies on the fourth Riemann
sheet. Interestingly, an analogous pole location was found in a lattice QCD
calculation of the $T$-matrix~\cite{Dudek:2016cru} who implemented a
coupled-channel generalisation of L\"{u}scher's single-channel
method~\cite{Luscher:1990ux}.  This result, however, corresponds to a pion
mass $\mpi=391$ MeV and it is not known how the pole location evolves upon
varying $\mpi$ (see, however, ref.~\cite{Guo:2016zep}). Fits to the Belle data
with a conventional $a_0(1450)$ were performed in ref.~\cite{Achasov:2010kk}
but only the integrated cross-sections were considered in that work. Detailed
global descriptions of photon-photon scattering to two mesons were proposed
also in ref.~\cite{Dai:2014zta} focusing mostly on the $I=0,2$ channels. In
the $I=1$ sector, they considered the $K\Kbar$ channel but not $\pi\eta$.

We perform here a global fit which takes into account both $\pi\eta$ and
$K\Kbar$ photon-photon data including all the differential cross-sections up
to 1.4 GeV.  In the $I=1$ sector, we use the $S$-wave coupled-channel
$T$-matrix model developed in ref.~\cite{Albaladejo:2015aca} which satisfies
unitarity, proper analyticity properties and matches to the chiral expansion
up to the next-to-leading order at low energy. The $S$-wave photon-photon
amplitudes are then deduced from a general MO representation involving two
subtraction constants and implementing a simple description of the left-hand
cut from cross-channel vector-meson exchanges. This is quite similar to
ref.~\cite{Danilkin:2017lyn}, we differ only by using $SU(2)$ chiral symmetry,
which allows to fix one of the subtraction constants through a soft pion
theorem\footnote{A further difference is that the soft photon
constraints at $s=0$ are not imposed in the dispersion relations used
in~\cite{Danilkin:2017lyn}.}.  
The $J=2$ partial-waves are described more
phenomenologically as a sum of cross-channel resonance exchange and a
direct $a_2(1320)$ Breit-Wigner amplitude. The $\gamma\gamma\to
(K\Kbar)_{I=1}$ amplitudes are then combined with $I=0$ amplitudes
taken from a previous work~\cite{GarciaMartin:2010cw} which considered
$\gamma\gamma\to (\pi\pi)_{I=0,2}, (K\Kbar)_{I=0}$ in order to
reconstruct the physical $\Kp\Km$, $\Kz\Kzb$ amplitudes. 
A global fit of photon-photon to meson-meson data was performed some time ago
based on unitarised chiral amplitudes~\cite{Oller:1997yg}, this model was also
applied to the $\eta\to \piz\gamma\gamma$
decay~\cite{Oset:2002sh,Oset:2008hp}. A remarkable qualitative agreement with
the data available at the time was achieved using a single arbitrary parameter
in the $S$-wave.  Today, a much larger data set is available and the precision
has increased significantly. We use here a model for the $T$-matrix involving
six parameters which will be determined by performing fits to these data.

An interesting aspect of the $\gamma\gamma\to \pi \eta$ amplitudes is that
they can be probed experimentally both in the scattering regime:
$s_{\gamma\gamma} \ge (\meta+\mpi)^2$ and in the decay regime: $0 \le
s_{\gamma\gamma} \le (\meta-\mpi)^2 $. In the decay region the amplitudes are
largely dominated by the light vector meson exchanges in the cross-channels:
$\gamma\pi \to \rho,\omega \to \gamma\eta$, which were first computed in
ref.~\cite{Ng:1992yg}. In this low-energy region the rescattering
contributions can be estimated in the $SU(3)$ chiral
expansion~\cite{Ametller:1991dp}. They proceed essentially via
$\gamma\gamma\to \Kp\Km \to \piz\eta$ but, at low energy, the $\pip\pim$
contribution (which is isospin violating) is not negligible and must also
be included~\cite{Ametller:1991dp}. We will present a dispersive calculation
of this contribution which gives rise to a cusp in the energy distribution
$d\Gamma^{\eta\to \piz \gamma\gamma}/ds_{\gamma\gamma}$.

The plan of the paper is as follows. After recalling some general
properties of photon-photon amplitudes and introducing the notation in
sec. 2, we write the unitarity relations for the partial-waves in
sec. 3 and present the modelling of the left-hand cut of these. In
sec. 4 we recall the derivation of a MO representation for the
$S$-waves. A dispersive representation for the isospin violating
$S$-wave, valid at low energy, is also derived. The comparison with
the experimental data on photon-photon scattering is performed in
sec. 5 and the information that can be deduced on the $a_0$
resonances are discussed.

\section{Basic ingredients}
\subsection{Kinematics}
We consider two-photon to two-meson scattering amplitudes
\be
\gamma(q_1)\gamma(q_2) \to M_1(p_1) M_2(p_2)
\en
with $M_1M_2= \pi\eta$ or $K\Kbar$. The Mandelstam variables are defined as
usual by
\be
s=(q_1+q_2)^2,\quad t=(q_1-p_1)^2,\quad u=(q_1-p_2)^2\ . 
\en
The various physical regions in the $s$, $t-u$ Mandelstam plane for the
$\gamma\gamma\to\pi\eta$, $\gamma\pi\to\gamma\eta$ and $\eta\to
\gamma\gamma\pi$ processes are shown on fig.~\fig{mandelstam}. 
In the centre-of-mass frame of the two photons the momenta are
expressed as
\be
\ba{ll}
q_1=\dfrac{\sqrt{s}}{2}\begin{pmatrix}
1\\
\hat{z}\\
\end{pmatrix},&\quad 
q_2=\dfrac{\sqrt{s}}{2}\begin{pmatrix}
1\\
-\hat{z}\\
\end{pmatrix} \\[0.4cm]
p_1=\dfrac{1}{2\sqrt{s}}\begin{pmatrix}
s+\Delta_{12}\\[0.2cm]
\sqrt{\lambda_{12}(s)}\,\hat{v}\\
\end{pmatrix},&\quad
p_2=\dfrac{1}{2\sqrt{s}}\begin{pmatrix}
s-\Delta_{12}\\[0.2cm]
-\sqrt{\lambda_{12}(s)}\,\hat{v}\\
\end{pmatrix}
\ea\en
where 
\be
\Delta_{12}=m_1^2-m_2^2,\quad 
\lambda_{12}(s)=(s-(m_1+m_2)^2)(s-(m_1-m_2)^2)\ .
\en
Taking $\hat{z}$ to a unit vector along the $z$ axis and $\hat{v}$ to be
a unit vector with polar angles $\theta$, $\phi$
such that $\hat{z}\cdot\hat{v}=\cos\theta$, we can express $t$, $u$ in
terms of $\cos\theta$ as,
\be\lbl{tuz}
\ba{l}
t=\dfrac{1}{2}\left(m_1^2+m_2^2-s + 
\sqrt{\lambda_{12}(s)}\cos\theta \right)\\[0.2cm]
u=\dfrac{1}{2}\left(m_1^2+m_2^2-s - 
\sqrt{\lambda_{12}(s)} \cos\theta\right)\ .
\ea
\en
\begin{figure}
\centering
\includegraphics[width=0.6\linewidth]{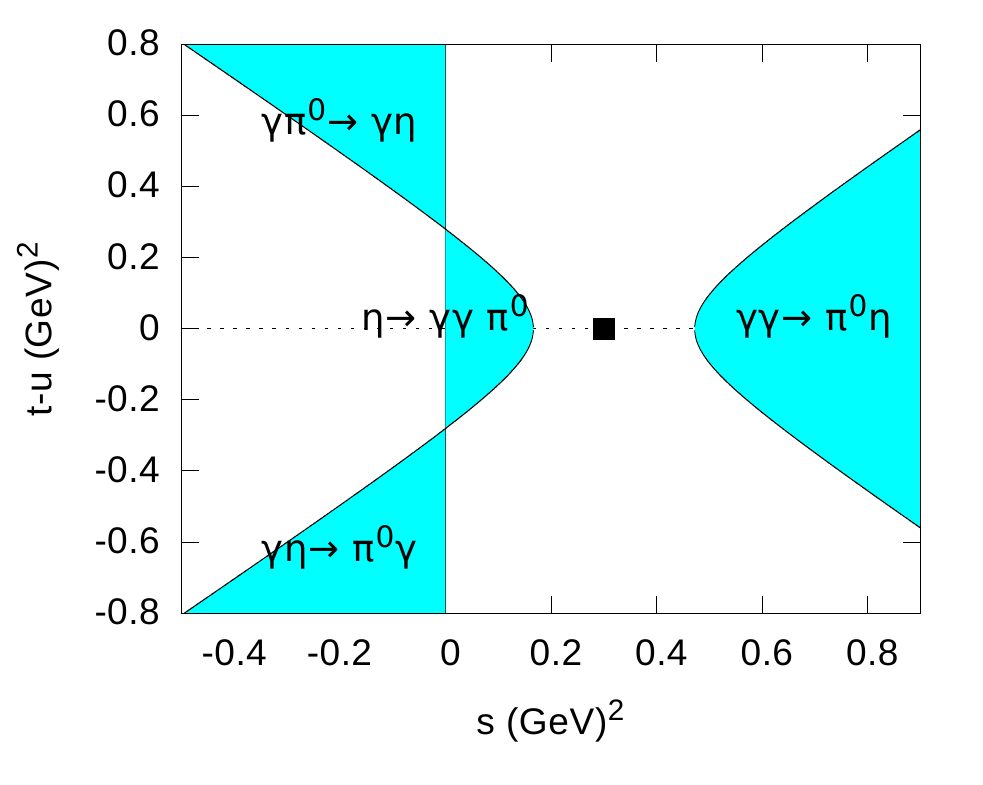}
\caption{\sl Physical regions for $\gamma\gamma\to \pi\eta$, $\gamma\pi\to
  \gamma\eta$ scattering and $\eta \to \gamma\gamma \pi^0$ decay. The black
  square indicates the soft pion point.}
\lblfig{mandelstam}
\end{figure}
The polarisation vectors of the two photons $\epsilon_1(q_1,\lambda)$, 
$\epsilon_2(q_2,\lambda')$ in this frame read
\be\lbl{eps1eps2}
\epsilon_1(q_1,\lambda)=\frac{1}{\sqrt2}\begin{pmatrix}
0 \\ -\lambda \\ -i \\ 0 \\
\end{pmatrix},\quad
\epsilon_2(q_2,\lambda')=\frac{1}{\sqrt2}\begin{pmatrix}
0 \\ \lambda' \\ -i \\ 0 \\
\end{pmatrix}
\en
and they satisfy
$\epsilon_1\cdot q_1 = \epsilon_2\cdot q_2 = \epsilon_1\cdot q_2=
\epsilon_2\cdot q_1=0\ .
$
\subsection{Amplitudes}
We denote the $\gamma\gamma\to \eta \pi$ helicity amplitude as
$L_{\lambda\lambda'}$ 
\be
\outleft\eta(p_1)\pi(p_2)\vert \gamma(q_1,\lambda)\gamma(q_2,\lambda')
\inright= ie^2 (2\pi)^4\delta(P_f-P_i)\,\hbox{e}^{i(\lambda-\lambda')\phi}
L_{\lambda\lambda'}(s,t)
\en
(we have factored out the electric charge $e$ and the dependence on
the azimuthal angle $\phi$)
while for the $\gamma\gamma\to K\Kbar$ amplitudes we use the same notation as
in previous work: $K^c_{\lambda\lambda'}(s,t)$ for charged kaons and
$K^n_{\lambda\lambda'}(s,t)$ for neutral kaons. Helicity amplitudes are
convenient for performing the partial wave expansion~\cite{Jacob:1959at}. They
can be expressed in terms of tensor amplitudes using the reduction
formulas, e.g.
\be
\hbox{e}^{i(\lambda-\lambda')\phi}L_{\lambda\lambda'}(s,t)=
\epsilon_1^\mu(q_1,\lambda)\epsilon_2^\nu(q_2,\lambda')
W_{\mu\nu}(q_1,q_2,p_1,p_2)\ . 
\en
By gauge invariance, the tensor amplitude $W_{\mu\nu}$ must satisfy the two
Ward identities
\be\lbl{wardid}
q_1^\mu W_{\mu\nu}= q_2^\nu W_{\mu\nu}=0\ .
\en
One can form two independent tensors which satisfy~\rf{wardid} which we take
as 
\be
\ba{l}
T_1^{\mu\nu}= \dfrac{1}{2} s g^{\mu\nu}-q_1^\nu q_2^\mu\\[0.2cm]
T_2^{\mu\nu}=
2s\Delta^\mu\Delta^\nu 
+4 q_1.\Delta\,q_2.\Delta\, g^{\mu\nu}
-4q_2.\Delta\, q_1^\nu\Delta^\mu
-4q_1.\Delta\, q_2^\mu\Delta^\nu
\ea\en
with
\be
\Delta=p_1-p_2 \ . 
\en
The tensor amplitude $W^{\mu\nu}$ can then be expressed in terms of two scalar
amplitudes 
$A$, $B$ 
\be
W^{\mu\nu}(q_1,q_2,p_1,p_2)= A(s,t,u) T_1^{\mu\nu} + B(s,t,u) T_2^{\mu\nu}
\en
which satisfy dispersion relations. Using eqs.~\rf{tuz}~\rf{eps1eps2} one can
easily express the helicity amplitudes $L_{\lambda\lambda'}$ in terms of the
two scalar amplitudes, 
\be
\ba{l}
L_{++}=L_{--}=\dfrac{s}{2}\,A(s,t)+s(2m_1^2+2m_2^2-s)\,B(s,t)\\[0.2cm]
L_{+-}=L_{-+}=\sin^2\theta \lambda_{12}(s) B(s,t)\ .\\ 
\ea\en
Assuming unpolarised photon beams the differential cross-section reads,
\be
\frac{d\sigma^{\gamma\gamma\to M_1M_2}}{d\cos\theta}=\frac{\pi \alpha^2}{4 s^2}
\sqrt{\lambda_{12}(s)} \left( \vert L_{++}\vert^2 + \vert
L_{+-}\vert^2\right)\ .
\en
Concerning the $\eta\to \pi\gamma\gamma$ decay amplitude, the double
differential distribution in the Dalitz plot reads,
\be
\frac{d^2\Gamma^{\eta\to\gamma\gamma \pi} }{ds dt}=\frac{\alpha^2}{8\pi m_\eta^3} 
 \left( \vert L_{++}\vert^2 + \vert L_{+-}\vert^2\right)\ 
\en
and the distribution as a function of $s$ only (which is the one available
experimentally) is given by
\be
\frac{d\Gamma^{\eta\to\gamma\gamma \pi} }{ds}
=\frac{\alpha^2}{32\pi m_\eta^3} \sqrt{\lambda_{12}(s)}
\int_{-1}^1 d\cos\theta\left( \vert L_{++}\vert^2 + \vert
L_{+-}\vert^2\right)\ .
\en
\subsection{Isospin}
Using the following (usual) isospin assignments for the pions and the
kaons 
\be
\begin{pmatrix}
\pip\\
\piz\\
\pim\\
\end{pmatrix}\sim
\begin{pmatrix}
-\ket{11}\\
\ket{10}\\
\ket{1,\!-\!1}\\
\end{pmatrix},\qquad 
\begin{pmatrix}
\Kp\\
\Kz\\
\end{pmatrix}\sim 
\begin{pmatrix}
\ket{\unun}\\[0.2cm]
\ket{\unmun}\\
\end{pmatrix},\qquad 
\begin{pmatrix}
\Kzb\\
\Km\\
\end{pmatrix}\sim 
\begin{pmatrix}
\ket{\unun}\\[0.2cm]
-\ket{\unmun}\\
\end{pmatrix}
\en
while $\eta\sim\ket{0,0}$, 
the relations between the amplitudes $\gamma\gamma\to \Kp\Km$, $\Kz\Kzb$
and the isospin amplitudes $\gamma\gamma\to (K \Kbar)_{I=0,1}$  read,
\be\lbl{Kisomatrix}
\left(\ba{l}
K^0_{\lambda\lambda'}\\
K^1_{\lambda\lambda'}
\ea\right)=
\left(\ba{rr}
-\sqrt{\frac{1}{2}} & -\sqrt{\frac{1}{2}}\\
-\sqrt{\frac{1}{2}} &  \sqrt{\frac{1}{2}}
\ea\right)
\left(\ba{l}
K^c_ {\lambda\lambda'}\\
K^n_ {\lambda\lambda'} 
\ea\right)\ , \en
and the analogous relations between the amplitudes
$\gamma\gamma\to  \pip\pim$, $\piz\piz$ and the corresponding isospin $I=0,2$
amplitudes (which will also be needed) is
\be
\left(\ba{l}
H^0_{\lambda\lambda'}\\
H^2_{\lambda\lambda'}
\ea\right)=
\left(\ba{rr}
-\sqrt{\frac{2}{3}} & -\sqrt{\frac{1}{3}}\\
-\sqrt{\frac{1}{3}} &  \sqrt{\frac{2}{3}}
\ea\right)
\left(\ba{r}
\sqrt2\,H^c_ {\lambda\lambda'}\\
H^n_ {\lambda\lambda'} 
\ea\right)\ .
\en
\section{Partial waves: unitarity, analyticity}
\subsection{Right-hand cut and unitarity relations}
It is convenient to collect the three $I=1$ scattering amplitudes $\pi\eta \to
\pi\eta$, $\pi\eta \to K\Kbar$ and $K\Kbar\to K\Kbar$ into a $2\times2$ matrix
\be
\bm{T}\equiv\begin{pmatrix}
T^{\pi\eta \to \pi\eta} & T^{\pi\eta \to K\Kbar}\\
T^{\pi\eta \to K\Kbar}  & T^{ K\Kbar \to K\Kbar}\\
\end{pmatrix}\en
and we can define the partial wave expansions as
\be
\bm{T}(s,t,u)=16\pi \sum (2j+1)\bm{T}_j(s) P_j(\cos\theta)
\en
where $\theta$ is the scattering angle in the centre-of-mass system. The
unitarity relation for the partial waves are easily derived and reads
\be
\im[\bm{T}_j(s)]= \bm{T}_j(s)\bm{\Sigma(s)} \bm{T}^*_j(s)
\en
with
\be
\bm{\Sigma(s)}=\begin{pmatrix}
\dfrac{ \sqrt{\lambda_{\pi\eta}(s)}}{s}\,\theta(s-(\mpi+\meta)^2) & 0\\
0 & \sqrt{\dfrac{s-4\mkd}{s}}\,\theta(s-4\mkd)\\
\end{pmatrix}\ .
\en
Concerning the $I=1$ photon-photon amplitudes we define the partial-wave
expansion as
\be\lbl{pwexpand}
\left(\ba{l}
L_{\lambda\lambda'}(s,t)\\
K^1_{\lambda\lambda'}(s,t)
\ea\right)=\sum_j (2j+1)
\left(\ba{l}
l_{j,\lambda\lambda'}(s)\\
k^1_{j,\lambda\lambda'}(s)
\ea\right)\,d^j_{\lambda-\lambda',0}(\theta)\ .
\en
The unitarity relations for the $S$-waves, as will be implemented below, read
\be\lbl{unitrelJ0}
\im\bp
l_{0++}(s)\\[0.3cm]
k^1_{0++}(s)\\
\ep=
\bm{T}^*_j(s)\bm{\Sigma(s)} \bp
l_{0++}(s)\\[0.3cm]
k^1_{0++}(s)\\
\ep\ ,
\en
which also give the discontinuities of the partial-waves (extended to complex
values of $s$) across the right-hand cut.
\subsection{$\pi\pi$ (isospin violating) contribution in $S$-wave unitarity}\label{sec:l0tilde}
The unitarity relation written above~\rf{unitrelJ0} collects the contributions
from the $\pi\eta$ and the $K\Kbar$ states.  Let us also consider the
contribution from the $\pi^+\pi^-$ state, which has the form
\be\lbl{pipiunitrel}
\im [\braque{\pi^0\eta\vert {\cal T} \vert \gamma\gamma}]_{\pi\pi}=
\frac{1}{2}\int d\Phi(p_+,p_-)\, 
\braque{\pi^0\eta\vert {\cal T} \vert\pip\pim}
\braque{\pip\pim\vert {\cal T}^\dagger \vert\gamma\gamma}
\en
where $d\Phi$ is the phase-space integration measure. This
contribution, being proportional to the $\pi\pi\to\eta\pi$ amplitude,
is isospin violating but it is enhanced at low energy due the large
size of the $\gamma\gamma\to \pi^+\pi^-$
amplitude~\cite{Ametller:1991dp}. The ChPT evaluation performed in
ref.~\cite{Ametller:1991dp} amounts to using the $O(p^2)$ tree-level
amplitudes for both $\pi\pi\to\eta\pi$ and $\gamma\gamma\to \pi^+\pi^-$ in
eq.~\rf{pipiunitrel}. An evaluation which goes beyond the chiral
expansion can be performed which we now discuss.

We consider only the $S$-wave contribution in
eq.~\rf{pipiunitrel} and a restricted kinematical region such that 
$s,t, u < 1$ $\hbox{GeV}^2$. In such a region, the $\pip\pim \to
\eta\piz$ amplitude can be approximated in terms of three one-variable
functions $M_0$, $M_1$, $M_2$
(see~\cite{Kambor:1995yc,Anisovich:1996tx}),
\be
\ba{l}
T^{\pip\pim \to \eta\piz}(s,t,u)=\\[0.2cm]
\qquad -\epsilon_L\big[M_0(s)-\frac{2}{3}M_2(s)
+(s-u)M_1(t)+(s-t)M_1(u)+M_2(t)+M_2(u)\big]\ .
\ea\en
An isospin violating parameter $\epsilon_L$ has been factorised which may be
taken as~\cite{Colangelo:2018jxw}
\be\lbl{epsilonL}
\epsilon_L=\dfrac{  \left({m}^2_\Kz- {m}^2_\Kp\right)_{QCD}}{3\sqrt3
  \fpid}\ .
\en
The three $M_I$ functions obey a set of coupled Khuri-Treiman
integral equations., see ref.~\cite{Colangelo:2018jxw} for a complete review of
work on this subject. We will use here the evaluation of the $\Kz-\Kp$
QCD mass difference from ref.~\cite{Colangelo:2016jmc} (updated
in~\cite{Colangelo:2018jxw}) based on experimental data on $\eta\to
3\pi$ decays: 
$\left({m}^2_\Kz-{m}^2_\Kp\right)_{QCD}=(6.24\pm0.38)\cdot10^{-3}$
GeV$^2$, which gives
\be
\epsilon_L=0.141\pm0.009\ .
\en
The amplitudes corresponding to a given  $\pi\pi$  isospin state $I,I_z$   
\be
{\cal M}^{I\,I_z}\equiv \braque{\eta\pi\vert T\vert \pi\pi;I\,I_z}
\en
are easily expressed using crossing symmetry and the Wigner-Eckart
theorem.  In the unitarity relation~\rf{pipiunitrel} the amplitudes
with $I=0,2$, ${\cal M}^{00}$ and ${\cal M}^{20}$, are needed which
have the following expressions
\be\lbl{MII_z}
\ba{l@{}r@{}l}
{\cal M}^{00}(s,t,u)=& \sqrt3\,\epsilon_L & \left[ M_0(s)+ 
  {1\over3}M_0(t)+{10\over9}M_2(t)+{2\over3}(s-u)M_1(t)
+(t\leftrightarrow u) \right] \\[0.2cm]
{\cal M}^{20}(s,t,u)=& -\frac{2\sqrt6}{3}\,\epsilon_L & \left[ M_2(s)+ 
  {1\over2}M_0(t)+{1\over6}M_2(t)-{1\over2}(s-u)M_1(t) 
+(t\leftrightarrow u) \right] \\
\ea\en
in terms of the $M_I$ functions. We denote the angular integrals of
these amplitudes as 
\be
\ba{l}
\frac{1}{2}\displaystyle\int_{-1}^1 dz\, {\cal M}^{00}(s,t,u)
\equiv\sqrt3\,\epsilon_L \left(M_0(s)+ \hat{M}_0(s)\right)\\[0.2cm]
\frac{1}{2}\displaystyle\int_{-1}^1 dz\, {\cal M}^{20}(s,t,u)
\equiv-\frac{2\sqrt6}{3}\,\epsilon_L \left(M_2(s)+\hat{M}_2(s)\right)\ .
\ea\en
With this notation, the $\pi\pi$ contribution to the unitarity
relation is finally expressed as follows,
\be\lbl{discl0tilde}
\ba{l@{}l}
\disc[{l}_{0,++}(s)]_{\pi\pi}=\epsilon_L
\dfrac{\sqrt3}{32\pi}\sqrt{\dfrac{s-4\mpid}{s}}
\,\bigg[ & \left(h^{0}_{0,++}(s)\right)^*(M_0(s)+\hat{M}_0(s))\\[0.2cm]
\ & -\frac{2\sqrt2}{3} \left(h^{2}_{0,++}(s)\right)^*(M_2(s)+\hat{M}_2(s))
\bigg]\\
\ea\en
where
\be
\disc[{l}_{0,++}(s)]_{\pi\pi}\equiv \frac{{l}_{0,++}(s+i\epsilon)
-{l}_{0,++}(s-i\epsilon)}{2i}\ ,
\en
and $h^I_{0,++}(s)$ are the two $S$-wave $\gamma\gamma\to (\pi\pi)^I$
amplitudes with $I=0,2$. This $\pi\pi$ discontinuity of ${l}_{0++}$
can be estimated from eq.~\rf{discl0tilde} using inputs from
ref.~\cite{GarciaMartin:2010cw} for $\gamma\gamma\to \pi\pi$ and from
~\cite{Albaladejo:2015aca} for $\pi\pi \to \eta\pi$. The result of
this estimate is illustrated on fig.~\fig{discl0tilde} and compared
with the chiral calculation at NLO. The dispersive evaluation displays a
square-root singularity at $s=(\meta-\mpi)^2$ induced by the endpoint
of the left-hand cut in the functions $\hat{M}_0$, $\hat{M}_2$ which
overlaps with the right-hand cut as a result of the instability of the
$\eta$. As a further consequence, the phase of the partial-waves
$M_I+\hat{M}_I$ violate Watson's theorem and do not cancel with the
phases of $h_{0,++}^I$ such that the discontinuity has both a real and
an imaginary part.
\begin{figure}
\centering
\includegraphics[width=9cm]{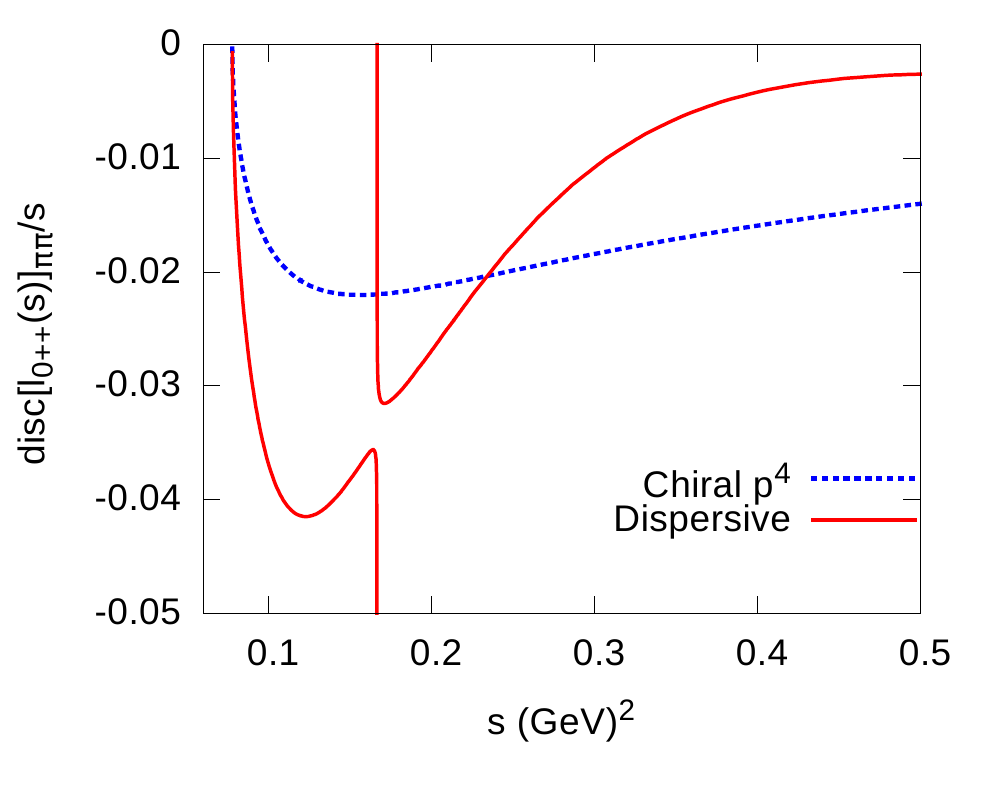}
\caption{\sl Discontinuity of the $\gamma\gamma\to \pi\eta$ $S$-wave
  amplitude (real part) across the $\pip\pim$ cut computed using dispersive
  results for the $\gamma\gamma\to \pi\pi$ and $\pi\pi\to \eta\pi$
  amplitudes, compared to the chiral $O(p^4)$ result.}
\lblfig{discl0tilde}
\end{figure}
\subsection{Left-hand cut: Born amplitudes}
A left hand cut in the $\gamma\gamma$ partial-waves is generated by
singularities in the cross-channels $\gamma P_1 \to \gamma P_2$. In
the case when $P_1=P_2=\Kp$ or $\pip$ the leading singularity is the
kaon or pion pole and the corresponding (so-called Born) amplitudes read
\be\lbl{BornAmplit}
\ba{l}
A_P^{Born}(s,t,u)=\dfrac{s}{(t-m^2_P)(u-m^2_P)}\\[0.3cm] 
B_P^{Born}(s,t,u)=\dfrac{1}{2(t-m^2_P)(u-m^2_P)}\\
\ea\en
with $P=\Kp,\ \pip$. We will need the $I=1$ component of the $K^+$
Born amplitude projected on the $S$-wave which reads,
\be\lbl{BornSproj}
k^{1,Born}_{0,++}(s)=-\frac{2\sqrt2\, m^2_{\Kp}}{s}\, L_\Kp(s),\quad
L_\Kp(s)=\frac{1}{\beta_{\Kp}(s)}\log\frac{1+\beta_{\Kp}(s)}
{1-\beta_{\Kp}(s)}
\en
with $\beta_P(s)=\sqrt{1-4m^2_P/s}$. We also recall the expressions of the
$J=2$ Born helicity amplitudes
\be
\ba{ll}
k^{1,Born}_{2++}(s)= & -\dfrac{2m^2_\Kp}{s\,\beta^2_\Kp(s)}\left[
( \beta^2_\Kp(s)-3)\,L_\Kp(s)+6  \right]      \\[0.4cm]
k^{1,Born}_{2+-}(s)= &  \dfrac{\sqrt6}{4s \beta^2_\Kp(s)}\left[
(1-\beta^2_\Kp(s))^2\,L_\Kp(s)+\dfrac{10}{3}\beta^2_\Kp(s)-2\right]\ .
\ea\en
\subsection{Left-hand cut: vector meson exchanges  }
Leading contributions to the left-hand cut in the $\piz\eta$ amplitudes
are modelled from the $\rho$, $\omega$, $\phi$ vector meson exchanges. 
We define the $VP\gamma$ coupling constants $G_{VP}$ through simple
Lagrangians 
 \be\lbl{lagLVg}
{\cal L}_{VP\gamma}= e G_{VP}\, \epsilon^{\mu\nu\alpha\beta} 
F_{\mu\nu}\partial_\alpha{P}V_\beta
\en
from which one can relate their values to the decay widths of the
vector mesons via 
\be
\Gamma_{V\to P\gamma}= \alpha C^2_{VP} \frac{ (m_V^2-m_P^2)^3}{6 m_V^3}\ .
\en 
The vector-exchange contributions to the $\gamma\gamma\to P_1 P_2$
amplitudes are easily computed,
\be
\ba{l}
A_V^{P_1P_2}(s,t,u)= G_{VP_1} G_{VP_2} \dfrac{-4t-2m_1^2-2m_2^2+s}{2(m_V^2-t)} 
+(t\leftrightarrow u)\\[0.25cm]
B_V^{P_1P_2}(s,t,u)= G_{VP_1} G_{VP_2} \dfrac{1}{4(m_V^2-t)}
+(t\leftrightarrow u)\ .\\ 
\ea\en
They give rise to poles in the zero-width approximation, which is
adequate in the kinematical regions of interest here.
The helicity amplitudes are
\be
\ba{l}
L_{V,++}^{P_1P_2}(s,\theta)=C_{VP_1}C_{VP_2}\dfrac{-s t}{m_V^2-t} 
+(t\leftrightarrow u)\\[0.3cm]
L_{V,+-}^{P_1P_2}(s,\theta)=C_{VP_1}C_{VP_2}
\dfrac{\sin^2\theta\lambda_{12}(s)}{4(m_V^2-t)}
+(t\leftrightarrow u)\\
\ea\en
and the corresponding partial-waves with $J=0,2$ have the following form
\be
\ba{ll}
l^V_{0,++}(s)=&  C_{V\pi}C_{V\eta}\, 2s\left(1- m_V^2 L_V(s) \right) \\[0.1cm]
l^V_{2,++}(s)=&   C_{V\pi}C_{V\eta}\, m_V^2 s \,\Big\{ (1-3X_V^2(s))L_V(s) 
+\dfrac{6X_V(s)}{\sqrt{\lambda_{12}(s)}}
\Big\}\\[0.3cm]
l^V_{2,+-}(s)=&   \frac{\sqrt6}{8}C_{V\pi}C_{V\eta}\,\Big\{
\lambda_{12}(s)(1-X_V^2(s))^2\,L_V(s) \\[0.2cm]
&\quad  -\frac{2}{3}\sqrt{\lambda_{12}(s)}X_V(s)(3X_V^2(s)-5)
\Big\}
\ea\en 
with
\be
X_V(s)=\frac{s-m_1^2-m_2^2+2m_V^2}{\sqrt{\lambda_{12}(s)}}\ 
\en
which is the cosine of the scattering angle when $t=\mvd$. The
function $L_V(s)$ is given by the angular integral
\be\lbl{Lvdef}
L_V(s)=\int_{-1}^1 dz\, \frac{s+2m_V^2- m_1^2-m_2^2}
{\lambda_{12}(s)(1-z^2) +4m_V^2(s\msv)}
\en
with
\be
s_V=-\frac{(m_V^2-m_1^2)(m_V^2-m_2^2)}{m_V^2}\ ,
\en
it can be expressed in terms of $X_V$ as
\be
L_V(s)=\frac{\log(X_V(s)+1)-\log(X_V(s)-1)}{\sqrt{\lambda_{12}(s)}}\ .
\en
We  note that the partial-wave $l^V_{0++}(s)$ has a soft
pion Adler zero at $s=s_A$ which can be approximated as
\be
s_A= \metad +\mpid\left(1 +\frac{\meta^2}{\mv^2}\left(
-\frac{2}{3}  
+\frac{4}{135}\frac{\mpi^2\meta^2}{\mv^4}
-\frac{8}{8505}\frac{\mpi^4\meta^4}{\mv^8}+\cdots\right)\right)\ .
\en

From the integral representation~\rf{Lvdef} one sees that the function
$L_V$ has  endpoint singularities at $z=\pm 1$ when $s=\psv$, which
thus corresponds to a branch point of $L_V(s)$. Another endpoint
singularity occurs when $s=\infty$. When $s > s_V$, the denominator
remains strictly positive. Therefore, $L_V$ is an analytic function of
$s$ with a cut on the negative real axis: $ -\infty < s < s_V$. The
discontinuity of $L_V$ along the cut is easily determined
\be
\im[L_V(s+i\epsilon)]=-\frac{\pi}{\sqrt{\lambda_{12}(s)}}\theta(s_V-s)
\en 
from which one deduces the left-cut discontinuities of the
vector-exchange partial-waves 
 \be\lbl{imVexchange}
\ba{ll}
\dfrac{1}{\pi}\im[l^V_{0,++}(s)]=&   2C_{V\pi}C_{V\eta}\, 
\dfrac{s\,m_V^2}{\sqrt{\lambda_{12}(s)}}\, \theta(s_V-s)    \\[0.4cm]
\dfrac{1}{\pi}\im[l^V_{2,++}(s)]=&  C_{V\pi}C_{V\eta}\,   
\dfrac{s\,  m_V^2}{\sqrt{\lambda_{12}(s)}} 
(3X_V^2(s)-1)\,\theta(s_V-s)    \\[0.4cm]
\dfrac{1}{\pi}\im[l^V_{2,+-}(s)]=&  -\dfrac{\sqrt6}{8}
C_{V\pi}C_{V\eta}\, \sqrt{\lambda_{12}(s)}\,
{(1-X_V^2(s))^2} \theta(s_V-s) \ .
\ea\en
We will use these discontinuities in the Muskhelishvili-Omn\`es
representations below.  We must consider also the vector exchange
contributions for the $\gamma\gamma\to (K\Kbar)_{I=1}$ amplitudes, we
denote the relevant combination of coupling constants as
\be
\tilde{C}_{K^*}^{(1)}\equiv \frac{1}{\sqrt2}\left(
-C^2_{K^*K^+}+ C^2_{K^*K^0}\right)\ .
\en
The imaginary parts of the $J=0,2$ partial-wave amplitudes along the
left-hand cut read
\be\lbl{imK*exchange}
\ba{ll}
\dfrac{1}{\pi}\im[k^V_{0,++}(s)]=&   2 \tilde{C}_{K^*}^{(1)}\, 
\dfrac{s\, m_{K^*}^2}{\sqrt{\lambda_{KK}(s)}} \theta(s_{K^*}-s)    \\[0.4cm]
\dfrac{1}{\pi}\im[k^V_{2,++}(s)]=&  \,\tilde{C}_{K^*}^{(1)}    
\dfrac{s\, m_{K^*}^2}{\sqrt{\lambda_{KK}(s)}} 
(3X_{K^*}^2(s)-1)\,\theta(s_{K^*}-s)    \\[0.4cm]
\dfrac{1}{\pi}\im[k^V_{2,+-}(s)]=&  -\dfrac{\sqrt6}{8}
\tilde{C}_{K^*}^{(1)}\, \sqrt{\lambda_{KK}(s)}\,
{(1-X_{K^*}^2(s))^2} \theta(s_{K^*}-s) \ .
\ea\en 
The updated values of the couplings $C_{VP}$ are collected in
table~\ref{tab:VectorResCP} below. 
\begin{table}[ht]
\centering
\begin{tabular}{cccccc}
\hline\hline
\TT\BB                  &$\Gamma$ (keV) &  $C_{VP}$(GeV$^{-1}$)  \\
\hline
$\rho^0\rightarrow\pi^0\gamma$    &$69(9)$   &  $0.368(24)$  \\
$\rho^0\rightarrow\eta\gamma$     &$44(3)$   &  $0.789(30)$  \\
$\omega\rightarrow\pi^0\gamma$    &$713(26)$ &  $1.160(20)$  \\
$\omega\rightarrow\eta\gamma$     &$3.8(4)$  &  $0.222(11)$  \\
$\phi\rightarrow\pi^0\gamma$      &$5.5(2)$   &  $0.067(1)$  \\
$\phi\rightarrow\eta\gamma$       &$55(1)$   &  $0.345(4)$  \\
$K^{*\pm}\rightarrow K^{\pm}\gamma$   &$50(5)$   &  $0.418(22)$  \\
$K^{*0}\rightarrow K^{0}\gamma$       &$116(11)$   &  $-0.636(30)$  \\
\hline\hline
\end{tabular}
\caption{\sl Radiative widths of vector mesons and corresponding coupling
  constants. The relative signs of the couplings are determined
  assuming flavour symmetry.}\label{tab:VectorResCP} 
\end{table}

\section{Representations of the $J=0,2$ partial-waves}

\subsection{Muskhelishvili-Omn\`es representations for the $S$-waves}
In order to write a dispersive representation for $l_{0++}$
some knowledge concerning its asymptotic behaviour is needed. Let us
then consider the angular integral,
\be
l_{0++}(s)\equiv \int_0^1 dz\, L_{++}(s,t)\ 
\en
in the $s\to \infty$ limit. There are two regions of the angular variable $z$ 
for which the behaviour of the integrand is known: a) when $z$ is
close to 0, then $t\sim u\sim -s\sim -\infty$. In this regime,
$L_{++}(s,t)$ can be estimated from QCD-based
methods~\cite{Brodsky:1981rp,Diehl:2001fv} according to which one
has
\be
 L_{++}(s,t) <<  L_{+-}(s,t) \sim \frac{\alpha_s(s)}{s}\ .
\en
b) When $z$ is close to 1, then $|t| << s$ which is the region where
Regge theory applies, the leading contribution from the vector meson
trajectory gives 
\be\lbl{regge} L_{++}(s,t) \sim \beta_V(t) (\alpha'
s)^{\alpha_V+\alpha' t}
\en
with $\alpha_V\simeq 0.5$, $\alpha'\simeq 0.9$ $\hbox{GeV}^{-1}$ and
$\beta_V(t)$ is a smooth function when $t<0$.  Assuming only that the
integrand evolves smoothly between these two regimes when $0\le z\le
1$, one deduces that the $l_{0++}(s)$ should not grow faster than
$\sqrt{s}$ when $s\to \infty$.

Furthermore, the $J=0$ partial-waves obey soft-photon theorems
~\cite{Low:1958sn,Abarbanel:1967wk} which imply that the ratios $l_{0++}(s)/s$
and $(k^1_{0++}(s)-k^{1,Born}_{0++}(s))/s$ remain finite when $s\to
0$. Therefore, they can be expressed as unsubtracted dispersion relations in
terms of the left-hand and right-hand cuts discontinuities,
\be\lbl{dispersive2g}
\ba{l}
l_{0++}(s)= s\left[
\dfrac{1}{\pi}{\displaystyle\int_{-\infty}^{\sbarv} }ds'\,
\dfrac{\im[l_{0++}(s')]}{s' (s'-s)} +
\dfrac{1}{\pi}{\displaystyle\int_{m_+^2}^\infty}ds'\,
\dfrac{\im[l_{0++}(s')]}{s' (s'-s)} \right] \\[0.4cm]
k^1_{0++}(s)=k_{0++}^{1,Born}(s)+s\left[
\dfrac{1}{\pi}{\displaystyle\int_{-\infty}^{\sbark} }ds'\,
\dfrac{\im[k^1_{0++}(s')]}{s' (s'-s)} +
\dfrac{1}{\pi}{\displaystyle\int_{m_+^2}^\infty}ds'\,
\dfrac{\im[k^1_{0++}(s')]}{s' (s'-s)}\right] 
\ea\en
with $m_+=m_\pi+m_\eta$. The isospin symmetry limit has been assumed here,
implying that the $\eta$ meson is stable and the discontinuities  are real.
At low energies the left-cut discontinuities are
dominated by the vector meson exchanges. We will introduce subtractions, in
order to reduce the influence of the higher energy regions where the
discontinuities are not known. The right-cut discontinuities are given by the
unitarity relations. Unitarity is known to be saturated with two channels to a
very good approximation in the region of the $a_0(980)$ resonance. We will
assume that elastic unitarity holds below the $K\Kbar$ threshold and that
two-channel unitarity remains a sufficiently good approximation up to
$\sqrt{s}=1.4$ GeV.  Under the assumption of two-channel unitarity the
dispersion relations~\rf{dispersive2g} form a set of coupled inhomogeneous
Muskhelishvili equations. They can be solved in terms of a two-channel MO
matrix which satisfies a homogeneous set of coupled equations in terms of the
$T$ matrix\footnote{The $\pi\eta$ and $K\Kbar$ scalar form factors are
  expected to go like $1/s$ at infinity (up to log's) in QCD and must be
  proportional to the matrix elements of the MO matrix (up to a
  polynomial). Consequently, the MO  matrix elements must vanish
  when $s\to\infty$ like $1/s$ at least, which is assumed in
  eq.~\rf{MOequations}.}
\be\lbl{MOequations}
\bm{\Omega}_0(s)=\dfrac{1}{\pi}\int_{m_+^2}^\infty \frac{ds'}{s'-s}
\bm{T}(s)\bm{\Sigma}(s)\bm{\Omega}^*_0(s)\ 
\en
Asymptotic conditions on the phase-shifts are imposed which ensure that 
the set of equations~\rf{MOequations} has a unique solution
once initial conditions at $s=0$ are given (see sec.~3.2 
in ref.~\cite{Albaladejo:2015aca} and references 
therein). At $s=0$ one can take
\be
\bm{\Omega}_0(0)=\begin{pmatrix}
1 & 0\\
0 & 1\\
\end{pmatrix}\ .
\en
Multiplying the amplitudes $(l_{0++},k_{0++})$ by the inverse of the matrix
$\bm{\Omega}_0$ removes the right-hand cuts. We can use this property in
writing once-subtracted dispersion relations for the two functions
\be
\bp
\phi_1(s)\\
\phi_2(s)\\
\ep\equiv
\bm{\Omega}_0^{-1}(s)
\bp
l_{0++}(s)/s\\[0.3cm]
(k^1_{0++}(s)- k_{0++}^{1,Born}(s))/s\\
\ep\ .
\en
This provides MO-type dispersive representations for the $\gamma\gamma$
amplitudes $l_{0++}(s)$,  $k_{0++}(s)$ in terms of their imaginary
parts on the left-cut and two parameters, $b_l$,  $b_k$
\be\lbl{omnesSwaves}
\bp
l_{0++}(s)\\
k^1_{0++}(s)\\
\ep= \bp
0\\
k_{0++}^{1,Born}(s)\\
\ep +
s\,\bm{\Omega}_0(s) \bp
b_l + L_1(s)+R_1(s) \\
b_k + L_2(s)+R_2(s) \\
\ep\ .
\en 
The functions $L_i(s)$ are dispersive integrals over the left-hand cuts.
We express them in a way which allows to easily implement the presence of an
Adler zero at $s=s_A$ in the amplitude $l_{0++}$ (see
appendix~\sect{chiralrev}) 
\be\lbl{dispSwaves}
\ba{ll} L_i(s)= &\dfrac{s-s_A}{\pi}
{\displaystyle\int_{-\infty}^\sbarv}\dfrac{ds'}{s'(s'-s_A)(s'-s)}
\,D_{i1}(s')\,\im[l^V_{0++}(s')]\\[0.4cm] \ & +\dfrac{s-s_A}{\pi}
{\displaystyle\int_{-\infty}^\sbark}\dfrac{ds'}{s'(s'-s_A)(s'-s)}
\,D_{i2}(s')\,\im[k^V_{0++}(s')]
\ea\en
where the functions $D_{ij}(s)$ are the matrix elements of the inverse
of the Omn\`es matrix,
\be
\bm{D}(s)\equiv \bm{\Omega}_0^{-1}(s)\ .
\en
The functions $R_i(s)$, secondly, are the dispersive integrals over the
right-hand cut,
\be
R_i(s)=-\frac{s-s_A}{\pi}\int_{4\mkd}^\infty 
\frac{ds'}{s'(s'-s_A)(s'-s)}\,
\im[D_{i2}(s')]\, k_{0++}^{1,Born}(s')\ .
\en
A relation between the parameters $b_l$
and $b_k$ can be derived from imposing that the amplitude $l_{0++}$ has an
Adler zero at $s=s_A$,
\be\lbl{Adlerblbk}
b_l= -b_k \Omega_{12}(s_A)/\Omega_{11}(s_A)\ .
\en
The parameter $b_k$ will eventually be fitted to the experimental data
but we can estimate its order of magnitude by matching the amplitude
$k^1_{0++}$ with the $SU(3)$ chiral expansion. Including the order $p^4$
contributions (see appendix\sect{chiralrev}) and an estimate of the
order $p^6$ from the vector-meson exchange amplitudes, we obtain
\be\lbl{b_kestimate}
\ba{ll}
b_k +L_2(0)+R_2(0) & \simeq 
-\dfrac{2\sqrt2}{\fpid}(L_9+L_{10})
-\sqrt2(G^2_{K^{*0}K}-G^2_{K^{*+}K})\dfrac{\mkd}{m_{K^*}^2}\\[0.3cm]
\                 &\simeq -(0.57\pm0.03)\ \hbox{GeV}^{-2}
\ea\en 
using the determination $L_9+L_{10}= (1.44\pm 0.08)\cdot10^{-3}$ taken from 
ref~\cite{Unterdorfer:2008zz}. Below, the value of $b_k$ will be fitted
to the experimental data, the resulting combination $b_k
+L_2(0)+R_2(0)$ turns out to have a sign compatible with~\rf{b_kestimate} and a
magnitude smaller by a factor of two. The result of the dispersive
construction of the $S$-wave amplitude $l_{0++}$ is illustrated in
fig.~\fig{l0dispersiv}. The corresponding result for the $K\Kbar$ amplitude
$k^1_{0++}$ is shown in appendix~\sect{amplitI=0} (see fig.~\fig{KKamplit}). 
\begin{figure}
\centering
\includegraphics[width=0.7\linewidth]{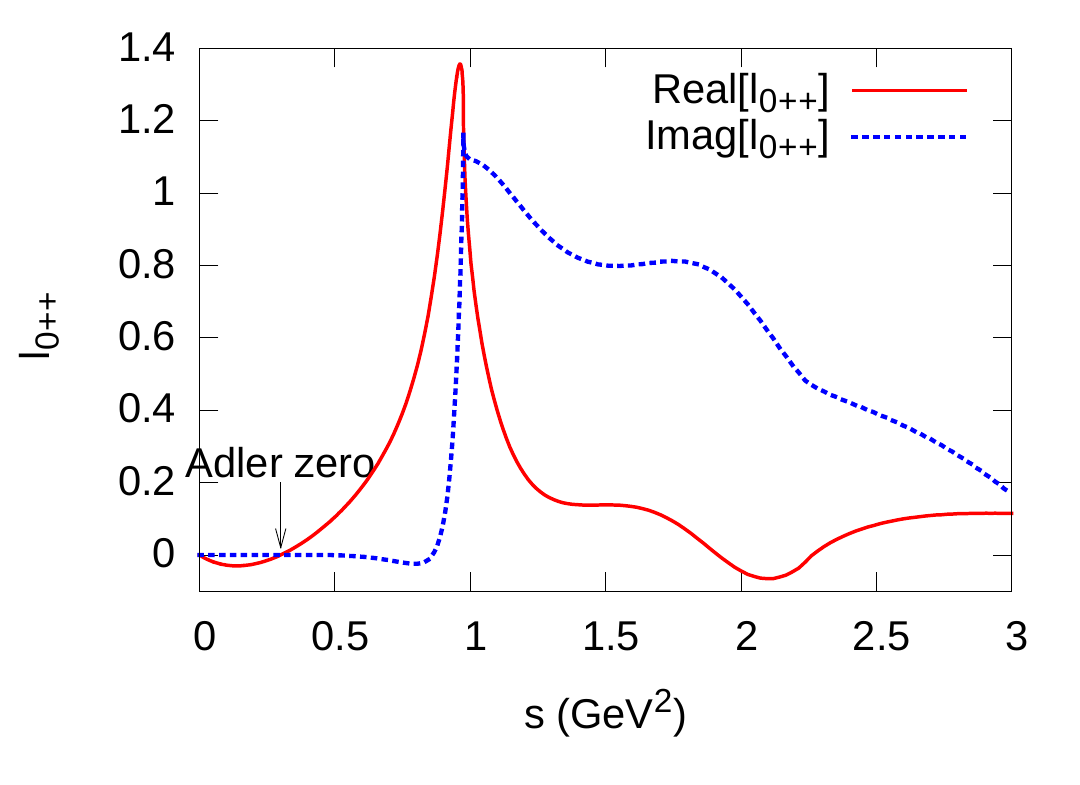}
\caption{\sl The $S$-wave amplitude $l_{0++}$ from the dispersive
  construction using the central values of the fitted parameters.}
\lblfig{l0dispersiv}
\end{figure}
\subsection{Dispersive construction of the isospin-violating $S$-wave}
In sec.~\sect{l0tilde} we have considered the unitarity contribution to the
$\gamma\gamma\to \pi\eta$ $S$-wave amplitude induced by the $\pip\pim$
intermediate state, which is isospin violating. Let us call $\tilde{L}_{++}$
the isospin-violating part of the $\gamma\gamma\to \pi\eta$ helicity amplitude
and $\tilde{l}_{0++}$ the corresponding $S$-wave. $\tilde{L}_{++}$ can be
defined as a matrix element,
\be\lbl{Ltildedef}
\tilde{L}_{++}\equiv
\braque{\piz(p_1)\eta(p_2)\vert \frac{1}{2}(m_d-m_u)(\bar{u}u-\bar{d}{d})
\vert\gamma(q_1,+)\gamma(q_2,+)}\ .
\en 
We will attempt here to estimate the amplitude $\tilde{l}_{0++}$ at low
energy only and we write a dispersive representation keeping only the
contribution from the $\pi\pi$ cut,
\be\lbl{l0tildeDR}
{\tilde{l}_{0,++}(s)}\equiv {s} \Big( \tilde{\lambda} + 
\dfrac{s-s_A}{\pi}{\displaystyle\int_{4\mpid}^\infty}
\dfrac{ds'}{s'(s'-s_A)(s'-s)}\,
\disc[{l}_{0,++}(s')]_{\pi\pi}  \Big)\ .
\en
We have used two subtractions in eq.~\rf{l0tildeDR} in order to strongly
reduce the influence of the energy region above 1 GeV. One subtraction
constant is fixed from imposing the soft photon zero. We can then estimate the
parameter $\tilde{\lambda}$ in eq.~\rf{l0tildeDR} by using the soft pion
limit. This limit provides a relation between the amplitude
$\tilde{l}_{0++}(s=s_A)$ and a matrix element of the pseudo-scalar
operator $p_0=i(\bar{u}\gamma^5u+\bar{d}\gamma^5d)$ which, in turn, can be
estimated using ChPT. This is detailed in
appendix~\sect{l0tildeapp}. Using eqs.~\rf{softpirel}~,\rf{softmatrix}
from this appendix we obtain the following result for
$\tilde{\lambda}$
\be
\tilde{\lambda}=
\dfrac{3\epsilon_L}{8\pi^2 s_A}\,\big(1-\dfrac{\mpid}{3\fpid}\big)
\,\big(G_\pi(s_A)-\frac{1}{2} G_K(s_A)\big)\ 
\en 
where the loop functions $G_\pi$, $G_K$ are given in eq.~\rf{G_P}.
The discontinuity  $\disc[{l}_{0++}(s)]_{\pi\pi}$ (given in
eq.~\rf{discl0tilde}) has a singularity at the pseudo-threshold
$s=(\meta-\mpi)^2$ induced by the $\pi\pi\to \eta\pi$ partial-wave (see
fig.~\fig{discl0tilde}).  The integral in eq.~\rf{l0tildeDR},
however, is finite. It is defined by using the $\metad+i\epsilon$ limiting
prescription exactly in the same way as those which appear in the
Khuri-Treiman equations for the $\eta\to 3\pi$ amplitude (see
e.g.~\cite{Kambor:1995yc,Gasser:2018qtg}). 

The result, in the low energy region relevant for $\eta \to \pi \gamma\gamma$
is shown in fig.~\fig{l0tildedisp} and compared to the corresponding chiral
$O(p^4)$ result from eq.~\rf{pi0etaNLO}. The chiral and dispersive real parts
agree at $s=\metad$, as a result of the soft pion relation, but the
two amplitudes differ substantially at lower energy:
\begin{itemize}
\item[$\bullet$]The cusp at the $\pi\pi$
threshold is much more pronounced in the dispersive amplitude which
is approximately five times larger in magnitude than the
$p^4$ amplitude at $s=4\mpid$. 

\item[$\bullet$] The $p^4$ amplitude has a zero at $s=4/3\mpid$ in contrast
to the dispersive amplitude which has no zero in this region. This is
because this zero is unrelated to a soft photon constraint and is an
accidental feature of the $p^4$ amplitude.
\end{itemize}
\begin{figure}
\centering
\includegraphics[width=10cm]{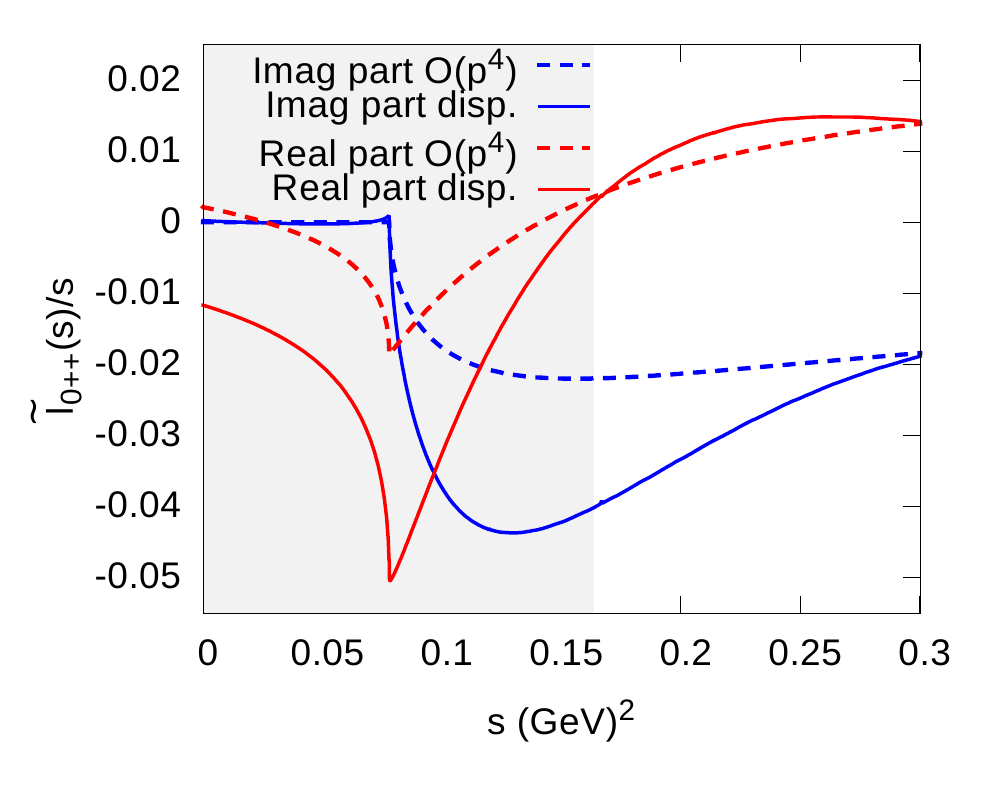}
\caption{\sl Dispersive calculation of the isospin-violating component of the
 $\eta\to \pi\gamma\gamma$  $S$-wave amplitude (divided by $s$), compared to the
  chiral result at order $p^4$. The shaded area shows the physical region.}
\lblfig{l0tildedisp}
\end{figure}

\subsection{$D$-wave amplitudes modelling:}
For the $J=2$ partial-waves one can write unsubtracted dispersion
representations analogous to those for $J=0$, e.g.,
\be\lbl{J=2disprels}
\ba{l@{}l}
l_{2\lambda\lambda'}(s)= s^{|\lambda+\lambda'|/2}\lambda_{12}(s)& \Big[
\dfrac{1}{\pi}{\displaystyle\int_{-\infty}^{\sbarv}} ds'\, 
\dfrac{\im[l_{2\lambda\lambda'}(s')]}
{(s')^{|\lambda+\lambda'|/2}\lambda_{12}(s')(s'-s)}\\[0.5cm]
\ & \qquad+\dfrac{1}{\pi}{\displaystyle\int_{m_+^2}^\infty} ds'\,
\dfrac{\im[l_{2\lambda\lambda'}(s')]}
{(s')^{|\lambda+\lambda'|/2}\lambda_{12}(s')(s'-s)} \Big]\\
\ea\en
displaying the kinematical zeros at $s=m_\pm^2$ and $s=0$. In the case of
$J=2$, however, it seems difficult to derive useful constraints from unitarity
as in the case of $J=0$ because in the important energy region of the
$a_2(1320)$ resonance there are too many contributing channels. We will
therefore content with a simple Breit-Wigner estimate of the right-hand cut
integral in eq.~\rf{J=2disprels}.

We parametrise the coupling of the $a_2$ resonance to the $\eta\pi$ channel
by a constant $C^{a_2}_{\eta\pi}$ defined from the Lagrangian
\be\lbl{LagTPP}
{\cal L}_{a_2\eta\pi}= C^{a_2}_{\eta\pi}\, T_{\mu\nu}(x) 
\partial^\mu\eta(x) \partial^\nu\pi^0(x)\ .
\en
The couplings to the $\gamma\gamma$ channel involve two constants
\be\lbl{LagTgg}
{\cal L}_{T\gamma\gamma}=e^2\,T_{\mu\nu}(x)\left\{
C^{a_2}_{\gamma\gamma}\, F^{\mu\beta}(x) F_\beta^{\phantom{\beta}\nu}(x)
+\frac{D^{a_2}_{\gamma\gamma}}{m_T^2}\, 
\partial^\mu F^{\alpha\beta}(x)\partial^\nu F_{\alpha\beta}(x)\right\}\ .
\en
The first term in~\rf{LagTgg} contributes to the $+-$ helicity state, which is
expected to be dominant, and the second term to the $++$ helicity state. The
expressions for the decay widths read,
\be\ba{l}
\Gamma[a_2 \to \eta\pi]=\dfrac{ (C^{a_2}_{\eta\pi})^2}{60\pi} 
\dfrac{ \left(q_{\eta\pi}(m_T^2)\right)^5}{m^2_T} \\[0.4cm]
\Gamma[a_2 \to \gamma\gamma]=\dfrac{e^4 m_T^3}{80\pi}\left(
(C^{a_2}_{\gamma\gamma})^2    +\dfrac{1}{6}(D^{a_2}_{\gamma\gamma})^2
\right)\ .
\ea\en
where $q_{\eta\pi}(s)=\sqrt{\lambda_{\eta\pi}(s)/4s}$. 
The experimental values of the branching fractions of the main hadronic decay
modes are~\cite{Tanabashi:2018oca} 
\be\ba{l}
B_{\eta\pi}=(14.5\pm1.2)\%,\quad
B_{K\Kbar}=(4.9\pm0.8)\%,\\
B_{3\pi}=(70.1\pm2.7)\%,\quad
B_{\omega\pi\pi}=(10.6\pm3.2)\%
\ea\en
and the $2\gamma$ width given by the PDG is 
\be
\Gamma^{a_2}_{\gamma\gamma}= 1.00\pm0.06\ \hbox{keV}\ .
\en
From this, one obtains the following values for the coupling constants
\be\lbl{CT_coupl}
\ba{ll}
C^{a_2}_{\eta\pi}=(10.8\pm0.5)\ \hbox{GeV}^{-1}\\[0.3cm]
\sqrt{(C^{a_2}_{\gamma\gamma})^2    +\dfrac{1}{6}(D^{a_2}_{\gamma\gamma})^2}=
(0.115\pm0.005)\ \hbox{GeV}^{-1}\
\ea\en
choosing $C^{a_2}_{\eta\pi}$ to have a positive sign.  Upon performing
fits to the differential cross-sections (see below) the two couplings
$C^{a_2}_{\gamma\gamma}$, $D^{a_2}_{\gamma\gamma}$ will get separately
determined. Defining a coupling constant $C^{a_2}_{K\Kbar}$ in the same 
way as $C^{a_2}_{\eta\pi}$ we get,
\be
C^{a_2}_{KK}=-(10.5\pm0.9)\ \hbox{GeV}\ 
\en
where the negative sign derives from flavour symmetry.
The Breit-Wigner model for the $a_2(1320)$ contributions is then taken as
\be\lbl{BWformulas}
\ba{l}
l_{2++}^{BW}(s')=\dfrac{D^{a_2}_{\gamma\gamma} C^{a_2}_{\eta\pi}}{60m_T^2}
\sqrt{\dfrac{W_2(q_{\eta\pi}(m_T^2)R)}{W_2(q_{\eta\pi}(s')R)}}
\dfrac{s'\lambda_{\eta\pi}(s')}{m_T^2-s'-im_T \Gamma_T(s)}\\[0.5cm]
l_{2+-}^{BW}(s')=\dfrac{\sqrt6\,C^{a_2}_{\gamma\gamma} C^{a_2}_{\eta\pi}}{60}
\sqrt{\dfrac{W_2(q_{\eta\pi}(m_T^2)R)}{W_2(q_{\eta\pi}(s')R)}}
\dfrac{\lambda_{\eta\pi}(s')}{m_T^2-s'-im_T \Gamma_T(s)}\ .
\ea\en
This form is obtained by first computing the amplitudes from the
Lagrangians~\rf{LagTPP} and~\rf{LagTgg} and then modifying them by introducing
a width function $\Gamma_T(s)$ in the denominator, for which we use the same
modelling\footnote{For two-body decay modes we take
  $\Gamma_{12}(s)=\Gamma^0_{12}\times (q_{12}(s)/q_{12}(s_0) )^{2l+1} W_l(
  q_{12}(s)R)/W_l( q_{12}(s_0)R)$ with $s_0=m_{a_2}^2$ and $l=2$ as in
  ~\cite{Uehara:2009cf} while for $\rho\pi$ we take $l=1$ and $W_1=1$. Such
  Breit-Wigner forms for tensor resonances are widely used but do not have
  good analyticity properties. In order to define the BW amplitudes below the
  thresholds we set the corresponding momenta to zero i.e. $q_{12}(s)=0$ if $s
  < (m_1+m_2)^2$.  }as in ref.~\cite{Uehara:2009cf}
\be
\Gamma_T(s)=\Gamma_{\eta\pi}(s)+\Gamma_{K\Kbar}(s)
+\Gamma_{\rho\pi}(s)+\Gamma_{\omega2\pi}(s)
\en
and a Blatt-Weisskopf related function (with $W_2(x)=9+3x^2+x^4$). Finally,
the $J=2$ amplitudes $l_{2\lambda\lambda'}$ are approximated by adding
the Breit-Wigner $a_2(1320)$ contribution in the $s$-channel to the
vector-exchange contributions in the $t$, $u$ channels,
\be
l_{2\lambda\lambda'}(s)= l^V_{2\lambda\lambda'}(s) +
l^{BW}_{2\lambda\lambda'}(s)\ .
\en

The corresponding $J=2$ $(K\Kbar)_{I=1}$ amplitudes are similarly described by
a sum of three terms, 
\be
k^1_{2\lambda\lambda'}(s)= 
-\frac{1}{\sqrt2}k^{Born}_{2\lambda\lambda'}(s) +k^{K^*}_{2\lambda\lambda'}(s) +
k^{BW}_{2\lambda\lambda'}(s)\ 
\en
where the Breit-Wigner amplitudes are given by
\be
\ba{l}
k_{2++}^{BW}(s')=\dfrac{D^{a_2}_{\gamma\gamma} C^{a_2}_{KK}}{60m_T^2}
\sqrt{\dfrac{W_2(q_{KK}(m_T^2)R)}{W_2(q_{KK}(s')R)}}
\dfrac{s\lambda_{KK}(s')}{m_T^2-s'-im_T \Gamma_T(s)}\\[0.5cm]
k_{2+-}^{BW}(s')=\dfrac{\sqrt6\,C^{a_2}_{\gamma\gamma} C^{a_2}_{KK}}{60}
\sqrt{\dfrac{W_2(q_{KK}(m_T^2)R)}{W_2(q_{KK}(s')R)}}
\dfrac{\lambda_{KK}(s')}{m_T^2-s'-im_T \Gamma_T(s)}\ .
\ea\en

It will be necessary to consider also the $(K\Kbar)_{I=0}$ amplitudes in order
to be able to construct the $\Kp\Km$ and $\Kz\Kzb$ amplitudes separately and
compare with the experimental results. From the value of the $K\Kbar$
branching fraction of the $f_2(1270)$ resonance~\cite{Tanabashi:2018oca}:
\be
B^{f_2}_{K\Kbar}=(4.6\pm0.5)\%\ ,
\en
one derives the following values for the corresponding coupling constants
\be\ba{l}
 C^{f_2}_{K\Kbar}=-(15.9\pm0.9)  \ \hbox{GeV}^{-1}\\[0.2cm]
\sqrt{ (C^{f_2}_{2\gamma})^2 + \frac{1}{6}(D^{f_2}_{2\gamma})^2}
=(0.19\pm 0.02) \ \hbox{GeV}^{-1}\ .\\
\ea\en
Based on nonet symmetry we have taken $C^{f_2}_{K\Kbar}$ and $C^{a_2}_{K\Kbar}$
to have the same sign.

\section{Comparison with experiment}
\subsection{Experimental inputs}
We will compare our model for the $\gamma\gamma$ amplitudes with precise
experimental data on $\gamma\gamma\to \pi\eta$ from the Belle
collaboration~\cite{Uehara:2009cf}, as was done recently in
ref.~\cite{Danilkin:2017lyn}. In addition, we consider also here
$\gamma\gamma\to K\Kbar$ data in order to provide further constraints on the
coupled-channel dynamics which is believed to be important for the $a_0(980)$
resonance. Recently, high statistics experimental data have been obtained by
the Belle collaboration for the $K_S K_S$ channel~\cite{Uehara:2013mbo}.
Experimental data for the charged kaons channel $K^+K^-$ are also available in
this low energy range ~\cite{Albrecht:1989re} but they are older and have much
less statistics.  We will restrict ourselves to the energy range $E\le 1.4$
GeV: we can use 448 differential cross-section points for $\pi\eta$ ($0.85 \le
E\le 1.39$ GeV) and 240 differential cross-section points for $K_SK_S$
($1.105\le E\le 1.395$ GeV).

\subsection{Parameters of the $T$-matrix}
Concerning the $S$-wave, firstly, we employ the $T$-matrix model of
ref.~\cite{Albaladejo:2015aca} which involves six parameters.  It uses a
chiral $K$-matrix type representation, which ensures one-channel unitarity
below the $K\Kbar$ threshold and two-channel unitarity above, together with
the chiral expansion. This kind of approach was initiated
in~\cite{Dobado:1989qm}, see ref.~\cite{Oller:2000ma} for a review. The
$T$-matrix is written as
\be\lbl{chiralK}
\bm{T}(s)=\left(\bm{1}- \bm{K}(s)\bm{\Phi}(s) \right)^{-1} \bm{K}(s)\
\en
where the matrix $\bm{\Phi}$ reads
\be\lbl{chiralK1}
\bm{\Phi}(s)=\bp
\alpha_1 +\beta_1 s +16\pi\bar{J}_{\eta\pi}(s) & 0 \\[0.2cm]
0 & \alpha_2 +\beta_2 s +16\pi\bar{J}_{KK}(s)
\ep\en 
it involves four phenomenological polynomial parameters $\alpha_i$, $\beta_i$
and the one-loop functions $\bar{J}_{P_1P_2}$ are given in~\rf{Jbardef}. The
$K$-matrix has the following form
\be\lbl{Kexpand}
\bm{K}(s)= \bm{K}_{(2)}(s) + \bm{K}_{(4)}(s) + \bm{K}_{(6)}(s)
\en
in which the subscript refers to the chiral order.
The first two terms are to be computed from the chiral
expansion of the scattering amplitudes involving the $\eta\pi$ and $K\Kbar$
channels at  order $p^2$ and $p^4$ respectively. The last term in
eq.~\rf{Kexpand} allows for a pole in $s$ and involves two phenomenological
parameters $m_8$ and $\lambda$,
\be\lbl{K6}
\left[\bm{K}_{(6)}(s)\right]_{ij}=\lambda \frac{g_i g_j}{16\pi}\left(
\frac{1}{m_8^2-s} - \frac{1}{m_8^2}
\right)\ ,
\en
The form of $g_1$, $g_2$ is derived from a resonance chiral
Lagrangian~\cite{Ecker:1988te}
\be\lbl{K6b}
\ba{l}
g_1=\dfrac{\sqrt6}{3\fpid}\left(c'_d(s-\metad-\mpid)+2c'_m \mpid\right)\\[0.3cm]
g_2=      \dfrac{1}{\fpid}\left(c'_d(s-2\mkd)+2c'_m\mkd\right)\ 
\ea\en
such that $\bm{K}_{(6)}$ has chiral order $p^6$ provided $\lambda$ is $O(1)$.
In this model, the $T$-matrix has good analyticity properties and
coincides with the chiral expansion at low energy up to $O(p^4)$
provided that the parameters $\alpha_i$, $\beta_i$, $\lambda$ are of
chiral order $O(1)$.  In addition to these six phenomenological
parameters, the $T$-matrix depends on the values of the $O(p^4)$
chiral parameters $L_i$~\cite{Gasser:1984gg} and on the ratio
$c'_m/c'_d$. We will use here the set of $L_i$ values from the $p^6$
fit of ref.~\cite{Bijnens:2014lea} (labelled as BE14 in that
reference). The ratio $c'_m/c'_d$ is expected to be of order $1-2$, we
will use $c'_m/c'_d=2$ as central value and include the variation as a source of
error.

Obviously, such a model which
implements two-channel unitarity is mostly justified in the $a_0(980)$ region
and below. We will assume that it remains qualitatively acceptable up to
$E\simeq 1.4$ GeV. The $T$-matrix is computed from eq.~\rf{chiralK} for $E \le
E_1$, $E_1=1.5$ GeV. In the higher energy region $E>E_1$, the $T$-matrix is
described through a simple interpolation of the phase-shifts and the
inelasticity such that $\delta_{11}(\infty)=2\pi$, $\delta_{22}(\infty)=0$ and
$\eta(\infty)=1$. These conditions introduce a smooth cutoff in the integral
equations satisfied by the matrix elements of the  MO matrix and ensure the
existence of a unique solution.

\subsection{Fits results}
In addition to the six $S$-wave $T$-matrix parameters listed above,
further parameters must be introduced which describe couplings to the
$\gamma\gamma$ channel.  In the $S$-wave, two parameters $b_l$, $b_k$
were introduced as subtraction constants. Implementing the Adler zero
condition we keep only $b_k$ as an independent parameter. In the
$D$-wave sector, we include the values of the tensor resonance
couplings $C^{a_2}_{2\gamma}$, $D^{a_2}_{2\gamma}$,
$C^{f_2}_{2\gamma}$, $D^{f_2}_{2\gamma}$ in the fitting as well as the
mass and width of the $a_2(1320)$ resonance: $m_{a_2}$,
$\Gamma_{a_2}$. In total, we thus have 6+7 parameters to be fitted.

At first, we have kept the $T$-matrix parameters fixed to one of the
sets of values determined previously in ref.~\cite{Albaladejo:2015aca}
(which, in particular, use assumed values for the pole positions of
the two $a_0$ resonances). It was not possible to obtain a good
fit of the $\gamma\gamma$ data in this manner: using these sets of parameter
values one finds that the $\pi\eta$ cross-section at the $a_0(980)$
peak tends to be too large and the energy of the peak tends to be
somewhat displaced as compared to experiment. 
Relaxing the $T$-matrix parameters, reasonably good fits become
possible and we actually found two distinct minimums of the total
$\chi^2$ combining the $\pi\eta$ and the $K_SK_S$ data,
\be
\ba{ll}
\chi^2\vert_{tot} &=(119+76)\vert_{\pi\eta} +233\vert_{K_SK_S} 
\qquad (\hbox{fit I})\\[0.2cm]
\      &=(93+117)\vert_{\pi\eta} +229\vert_{K_SK_S} 
\qquad (\hbox{fit II})\ .
\ea\en
In the case of $\pi\eta$ the first number corresponds to the
region $E<1.1$ GeV. In this region fit II is better than fit I while
the overall $\chi^2$ is slightly smaller in fit I ($\chi^2=428$) than
in fit II ($\chi^2=439$).  The $\chi^2$ is defined in a simple and
naive way: the correlation matrix is assumed to be diagonal and the
statistical and systematic errors provided by the Belle collaboration
are added in quadrature. The searches for minimums were performed with
the help of the computer code MINUIT~\cite{James:1975dr}. 

\begin{table}[h]
\centering
\begin{tabular}{c|c|c|c|c|c|c}
\hline\hline
\TT \BB & $\alpha_1$ & $\beta_1(\hbox{GeV}^{-2})$ & 
$\alpha_2$ & $\beta_2(\hbox{GeV}^{-2})$ & $m_8(\hbox{GeV})$&$\lambda$ \\ \hline
\TT fit I & $4.00(8)$& $-2.23(4)$ & $-0.545(5)$ & $0.167(6)$ & $1.304(4)$ & $0.47(4)$\\[0.2cm] 
fit II& $0.98(3)$& $-4.07(1)$ & $-0.495(1)$ & $-0.18(1)$ & $0.900(2)$ & $1.064(1)$\\
\hline\hline
\end{tabular}
\caption{\sl Parameters of the two-channel $T$-matrix in the two
  fits.}\label{tab:Tmatcoupl}  
\end{table}
The numerical values of the set of $T$-matrix parameters resulting from
these two fits are shown in table~\ref{tab:Tmatcoupl}. One notices, in
particular, that the value of the pole parameter $m_8$ differs significantly
in the two fits. Fig.~\fig{compphas} shows the behaviour of the two
phase-shifts $\delta_{11}$, $\delta_{22}$ and of the inelasticity parameters
$\eta$ as a function of energy, corresponding to the two fits.  At low
energies, the $\pi\eta$ phase-shift changes sign and becomes negative. This
low-energy behaviour is not anticipated in simple hadronic models of the
$\pi\eta$ amplitude (e.g.~\cite{Black:1999dx}) but it was also observed to
emerge from fitting the $\gamma\gamma$ data in
ref.~\cite{Achasov:2010kk}. 

%
In fit II the $\pi\eta$ scattering length (defined as in
~\cite{Bernard:1991xb}) is found to have the following value
\be
\mpi a_0^{\pi\eta}= -(9.5^{+0.3}_{-6.3})\times 10^{-3}\ .
\en
For comparison, the first two terms in the chiral expansion of the scattering
length give
\be
\mpi a_0^{\pi\eta}= \left.6.17\times 10^{-3}\right\vert_{p^2},\ 
=\left.2.43\times 10^{-3}\right\vert_{p^2+p^4}                       \ .
\en
using the same set of chiral parameters as in the unitary $T$-matrix.
\begin{figure}
\centering
\includegraphics[width=0.49\linewidth]{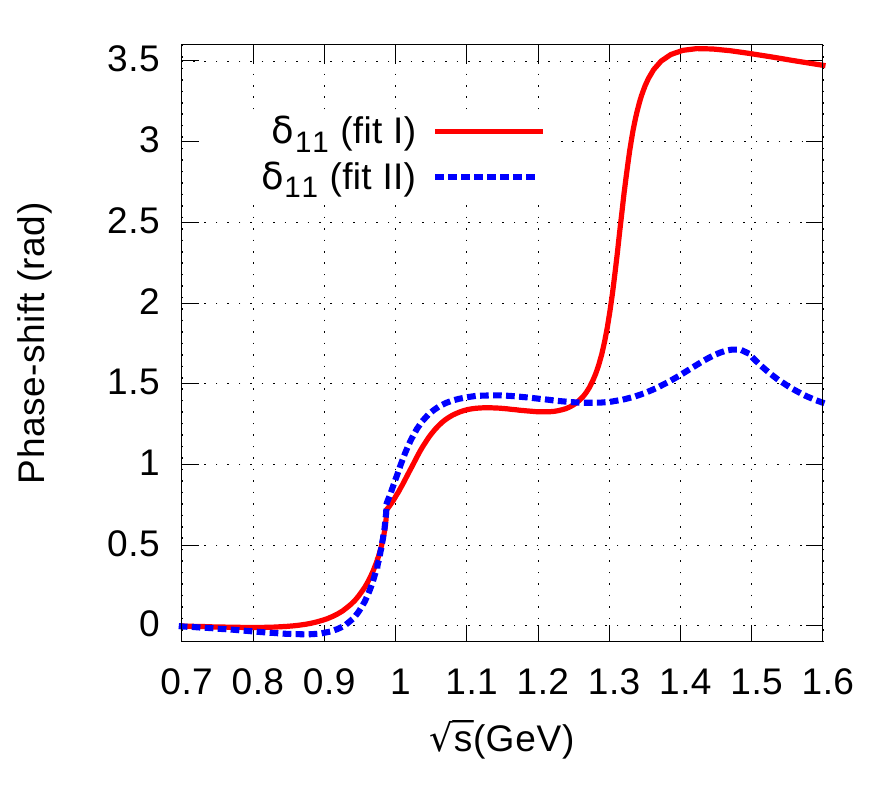}\hfill\includegraphics[width=0.49\linewidth]{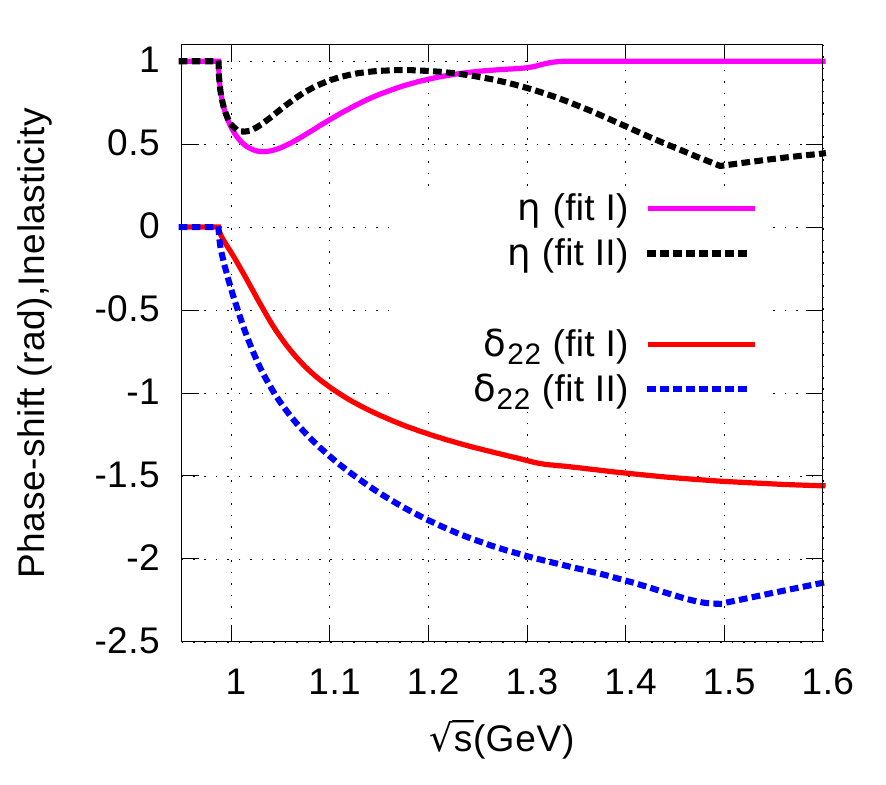}
\caption{The phase-shifts and the inelasticity from the two-channel $T$-matrix
model using the two sets of parameters corresponding to the two $\chi^2$ minimums.}
\lblfig{compphas}
\end{figure}

At higher energies a clear difference between the two fits is the
sharp increase of the $\pi\eta$ phase-shift in fit I around
$E\simeq1.32$ GeV, typical of a narrow resonance, which we will call
$a'_0$. Indeed, one finds a resonance pole in the $T$-matrix located
on the third Riemann sheet with the value\footnote{ The errors quoted
  here are the statistical ones as evaluated by MINUIT. Further errors
  introduced by varying intrinsic parameters of the model will be
  considered below.},
\be
\sqrt{s_{a'_0}}=1315(4)-i\,24(3)\ \hbox{MeV}\qquad (\hbox{fit I})\ .
\en
This resonance is lighter and narrower than the standard $a_0(1450)$.  Since
the phase-shift increases from $\pi/2$ to $\pi$ (approximately) it gives rise
to a sharp dip (instead of a peak) in the $S$-wave cross-section. Clearly
then, our fit I is quite analogous to the best fit by the Belle
collaboration~\cite{Uehara:2009cf} which displays a resonance (called
$a_0(Y)$ in that reference) which has very similar features while using a
parametrisation rather different from ours\footnote{The Belle collaboration
  parametrise the $S$-wave amplitude $l_{0++}$ as a sum of two Breit-Wigner
  functions plus a polynomial quadratic in the energy $E$. This representation
  involves 11 free parameters.}.
The $S$-wave amplitude in fit II also has an $a'_0$ resonance pole but
it is heavier and much broader than that of fit I, 
\be\lbl{a01450pole}
\sqrt{s_{a'_0}}=1421(5)-i\,175(4)\ \hbox{MeV}\qquad (\hbox{fit II})\ .
\en
In this case, the width of the resonance is larger than the experimental one.
One eventually does not expect a very accurate determination since the
resonance lies close to the cutoff of the model.

\begin{figure}
\centering
\includegraphics[width=0.333\linewidth]{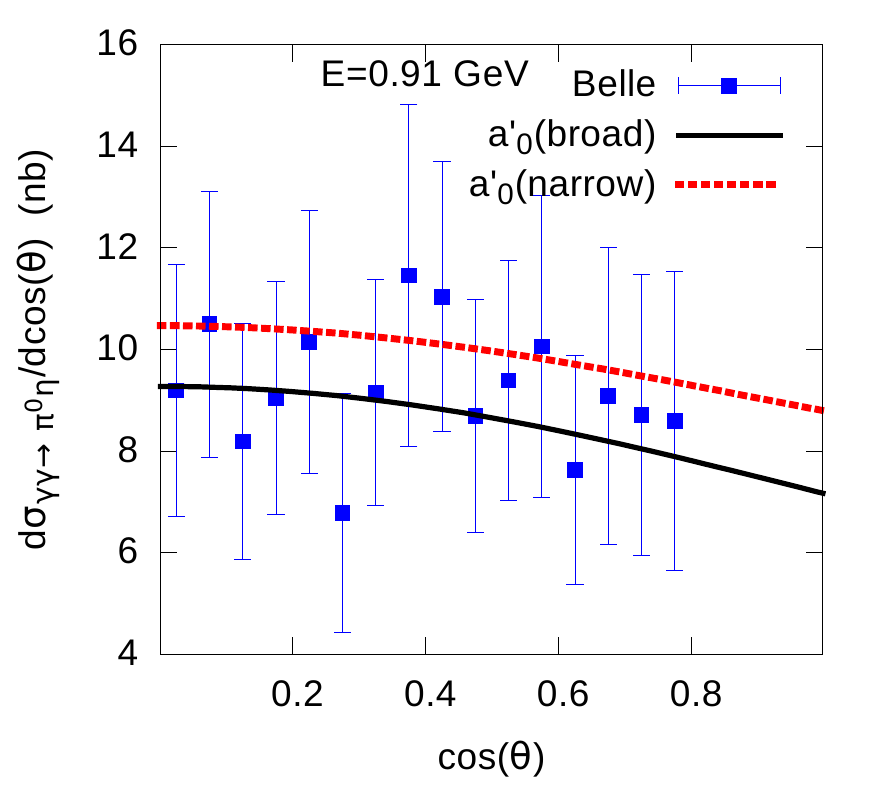}\includegraphics[width=0.333\linewidth]{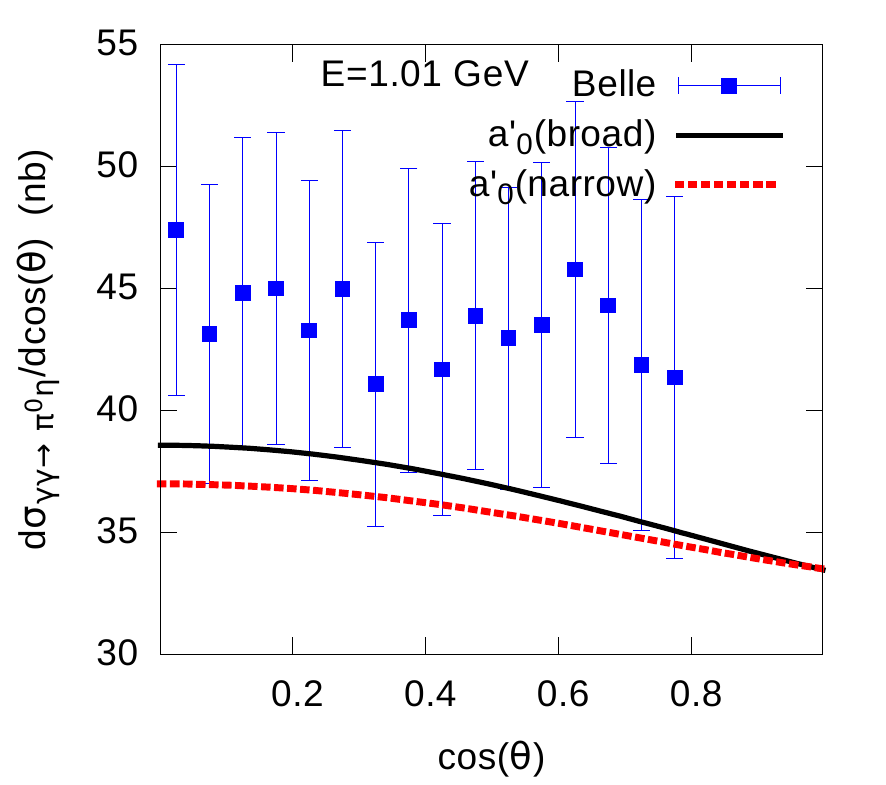}\includegraphics[width=0.333\linewidth]{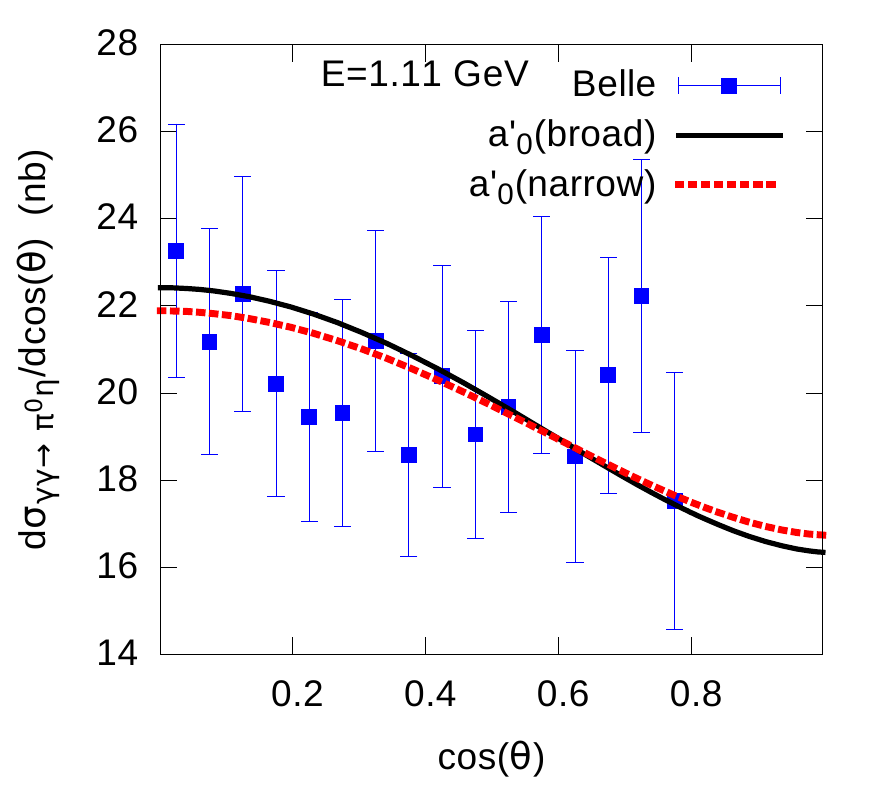}\\
\includegraphics[width=0.333\linewidth]{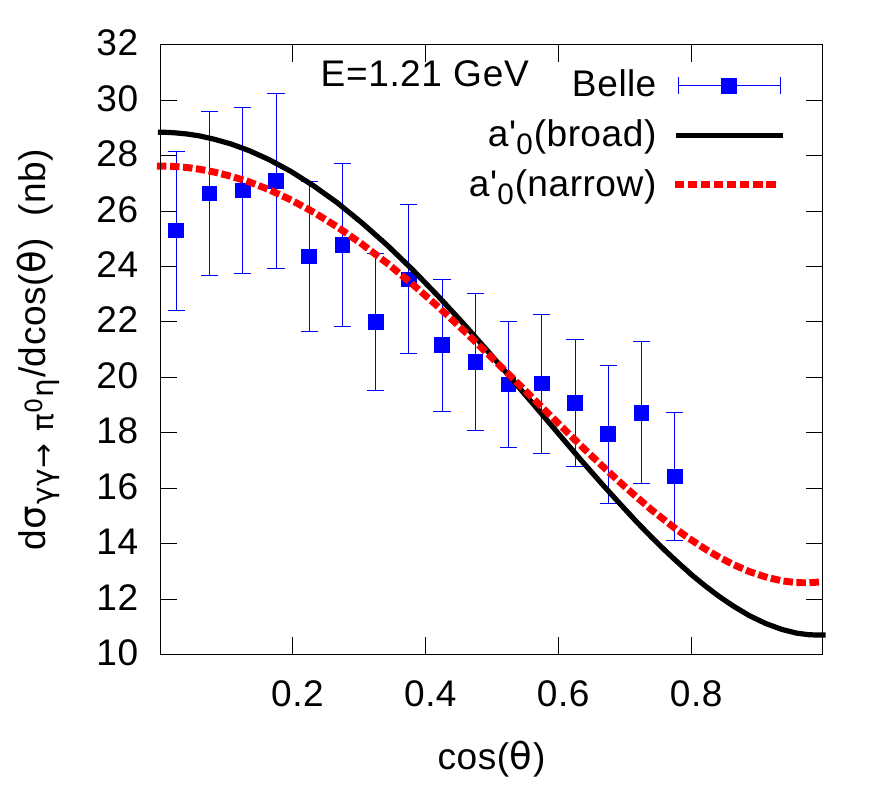}\includegraphics[width=0.333\linewidth]{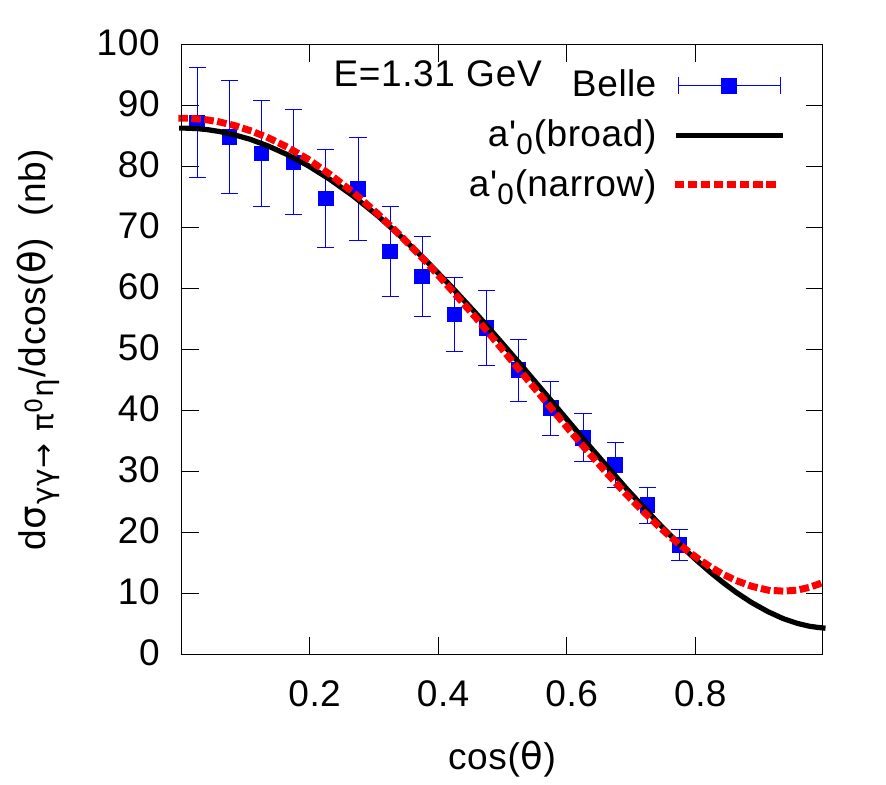}\includegraphics[width=0.333\linewidth]{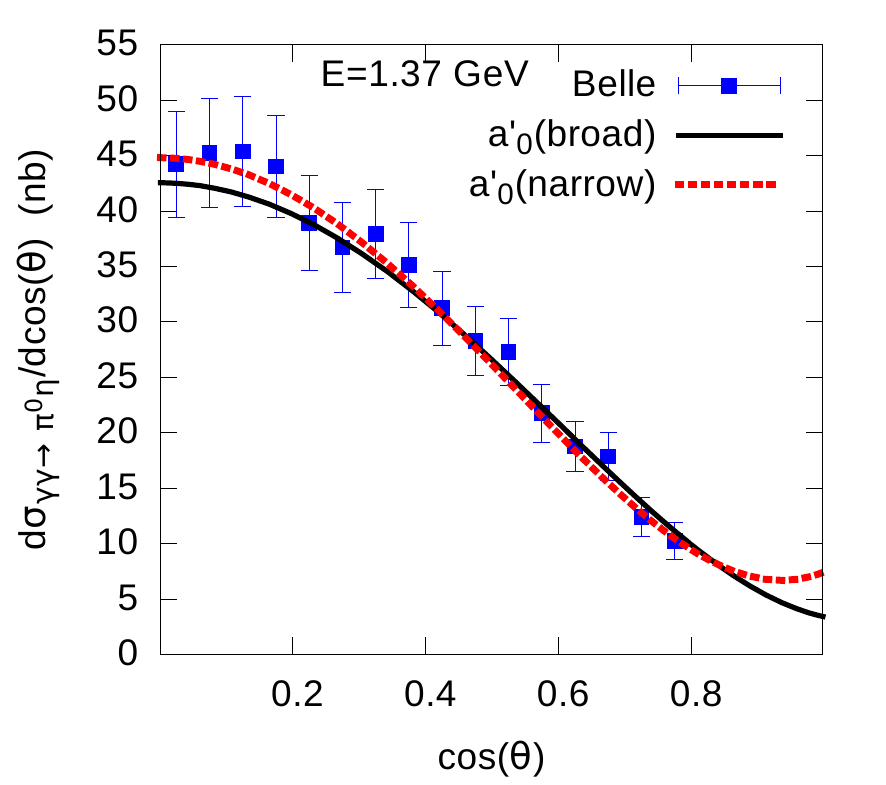}\\
\caption{\sl Experimental $\gamma\gamma\to \pi\eta$  differential
  cross-sections compared with the two fits results.}
\lblfig{diffcrosspieta}
\end{figure}

Fig.~\fig{diffcrosspieta} and fig.~\fig{diffcrossKsKs} show a sample of
$\gamma\gamma\to \pi\eta, K_SK_S$ differential cross sections comparing the
experimental results with those from the two fits. The difference between the
two fits is remarkably small, which is somewhat puzzling: why is the narrow
$a'_0$ resonance present in fit I not seen much more clearly?  The reason for
this arises from a specific interference effect between the
$J^{\lambda\lambda'}=0^{++}$ and the $2^{++}$ amplitudes. We have seen already
that the fast energy variation induced by the narrow $a'_0$ coincides with
that induced by the $a_2(1320)$.  More specifically, when $\sqrt{s}\simeq 1.32$
GeV, the following relations hold approximately between the $l_{0++}$ and the
$l_{2++}$ amplitudes in the case of fit I
\be
\ba{l}
\re[l_{0++}(s)] \simeq -5\re[l_{2++}(s)]\\[0.2cm]
\im[l_{0++}(s)] \simeq -5\im[l_{2++}(s)]+0.7\ .
\ea\en
The sum of the $0^{++}$ and the $2^{++}$ partial-wave amplitudes can
effectively be absorbed into the $2^{+-}$ resonance amplitude (since the
angular functions satisfy the relation $d^2_{00}(\theta)-1 = -\sqrt6\,
d^2_{20}(\theta)$) and  no specific $0^{++}$ resonance effect remains. We
also note that the $a'_0$ in fit I essentially decouples from the $K\Kbar$
channel such that no $0^{++}$ resonance effect is seen in this channel
either. Because of these fine-tuned relations between the scalar and the
tensor resonances the fit I solution is most likely to be unphysical.

\begin{figure}
\centering
\includegraphics[width=0.5\linewidth]{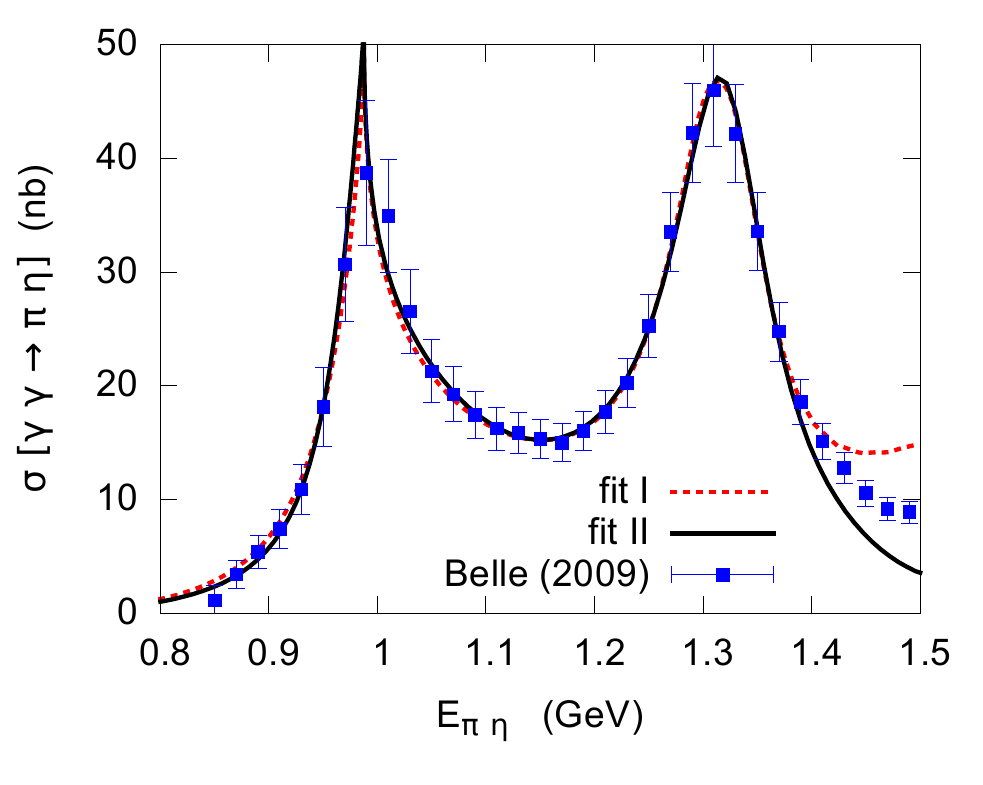}\includegraphics[width=0.5\linewidth]{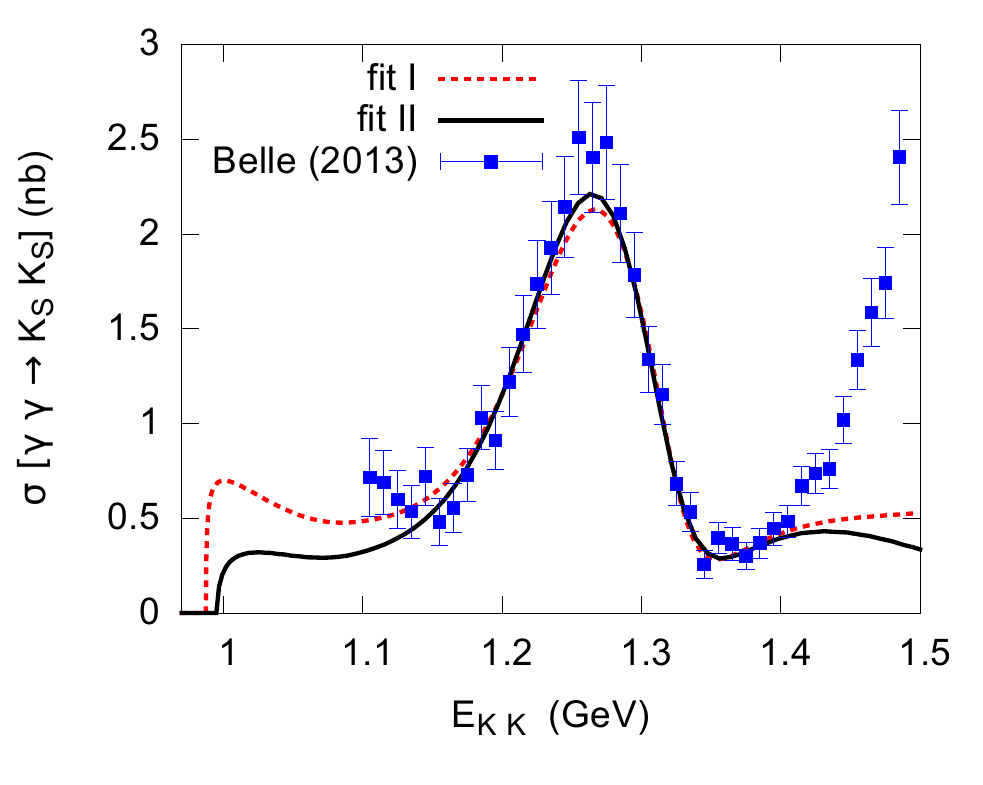}\\
\includegraphics[width=0.5\linewidth]{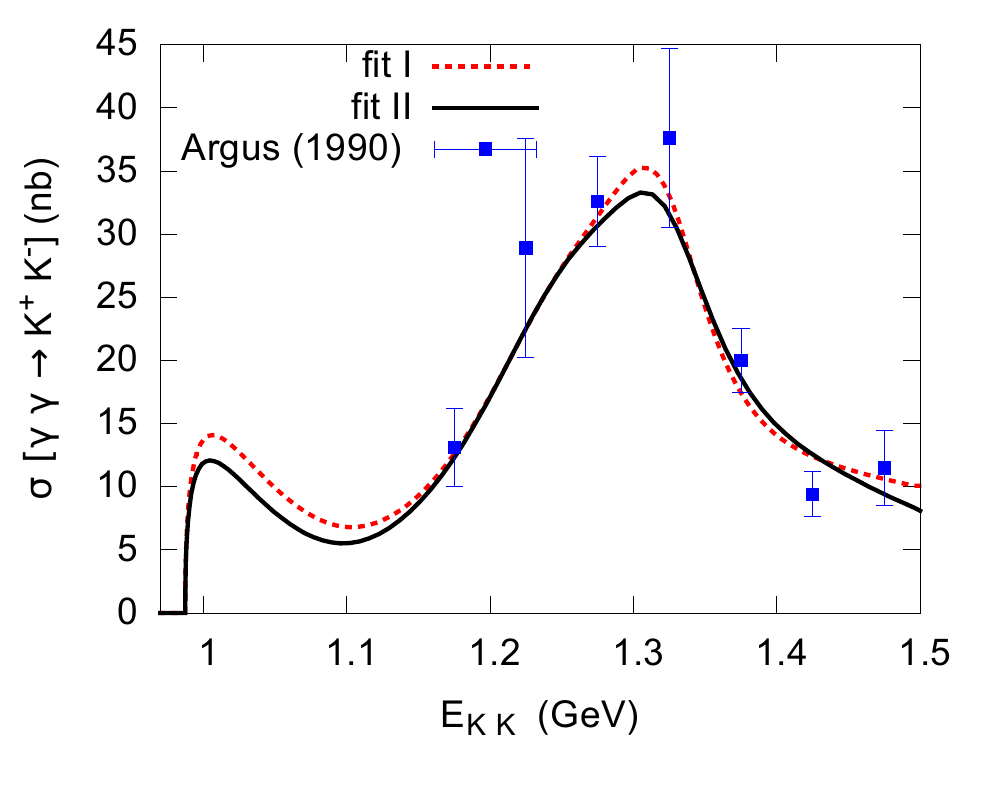}
\caption{\sl Cross-sections for $\gamma\gamma\to \pi\eta, K_SK_S,
  \Kp\Km$ integrated in the range $|\cos\theta|<0.8$. The data are
  from refs.~\cite{Uehara:2009cf,Uehara:2013mbo,Albrecht:1989re}, they
are compared with the two fits results.}
\lblfig{integcross}
\end{figure}
One can see from fig.~\fig{diffcrosspieta} that the $\pi\eta$ differential
cross-sections are well described except, however, in one energy bin: $E=1.01$
MeV. At this energy, the cross-sections are underestimated by our model, see
also fig.~\fig{integcross} which shows the cross-sections integrated
over $\theta$. 
This discrepancy could be caused by an isospin breaking effect close to the
resonance peak. Our model assumes isospin symmetry and uses $m_K\equiv m_\Kp$
(in order to correctly implement the kaon Born amplitudes) and the peak of the
cross-section occurs exactly at $E=2m_\Kp$. Physically, however, the $K^+$ and
the $K^0$ have slightly different masses which should lead to two cusps in the
shape of the integrated cross-section. On the experimental side, the energy
resolution should be smaller than $2m_\Kp-2m_\Kz\simeq8$ MeV in order to
clearly observe this effect.

\begin{figure}
\centering
\includegraphics[width=0.333\linewidth]{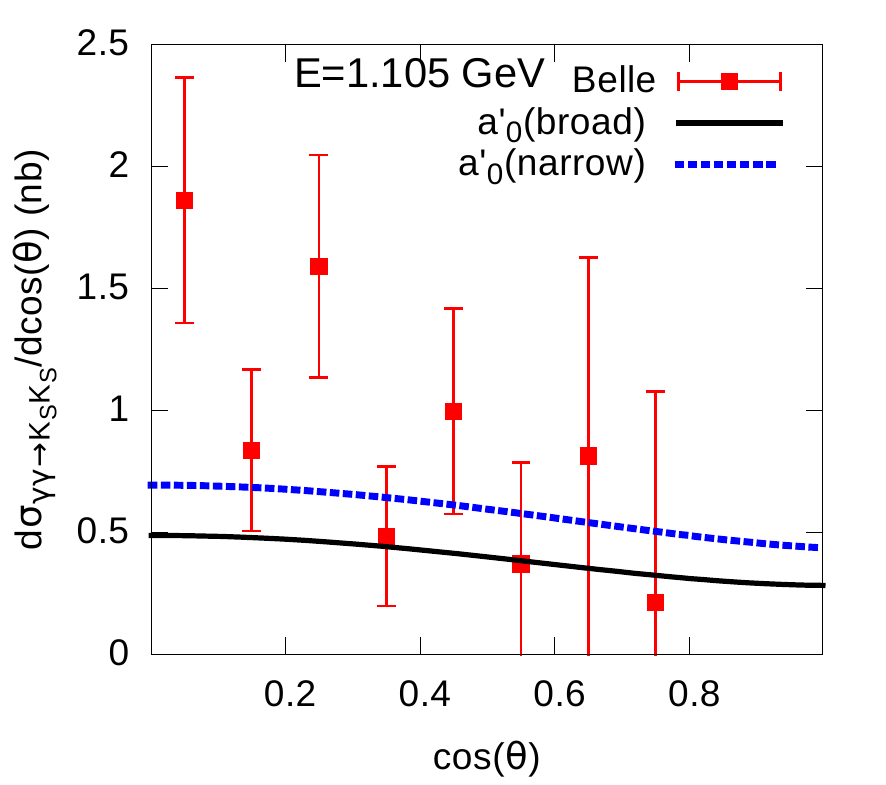}\includegraphics[width=0.333\linewidth]{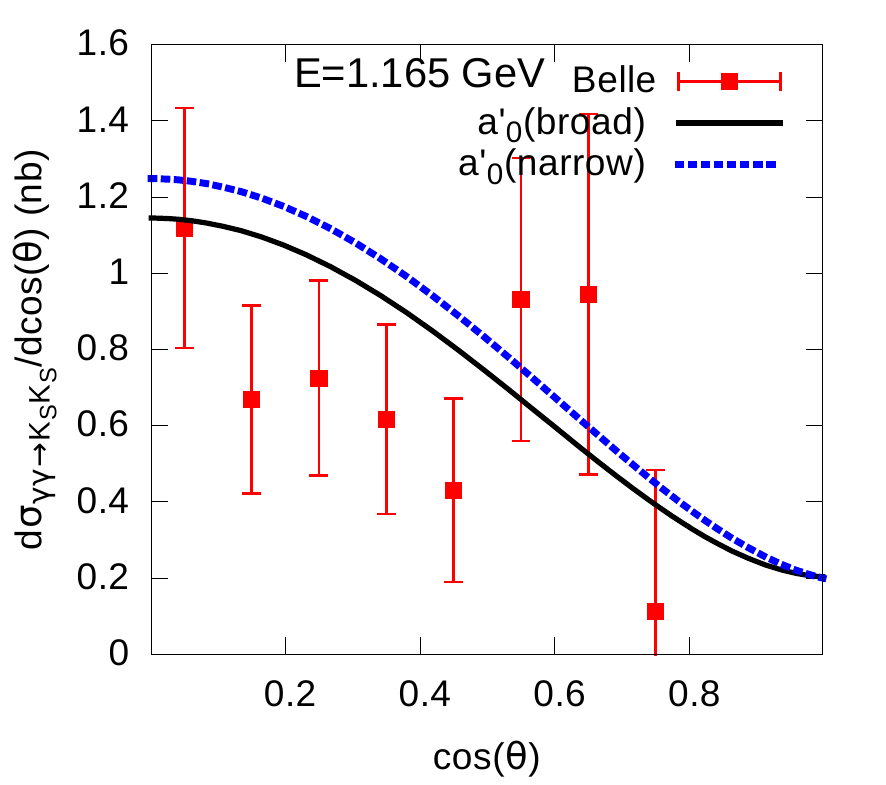}\includegraphics[width=0.333\linewidth]{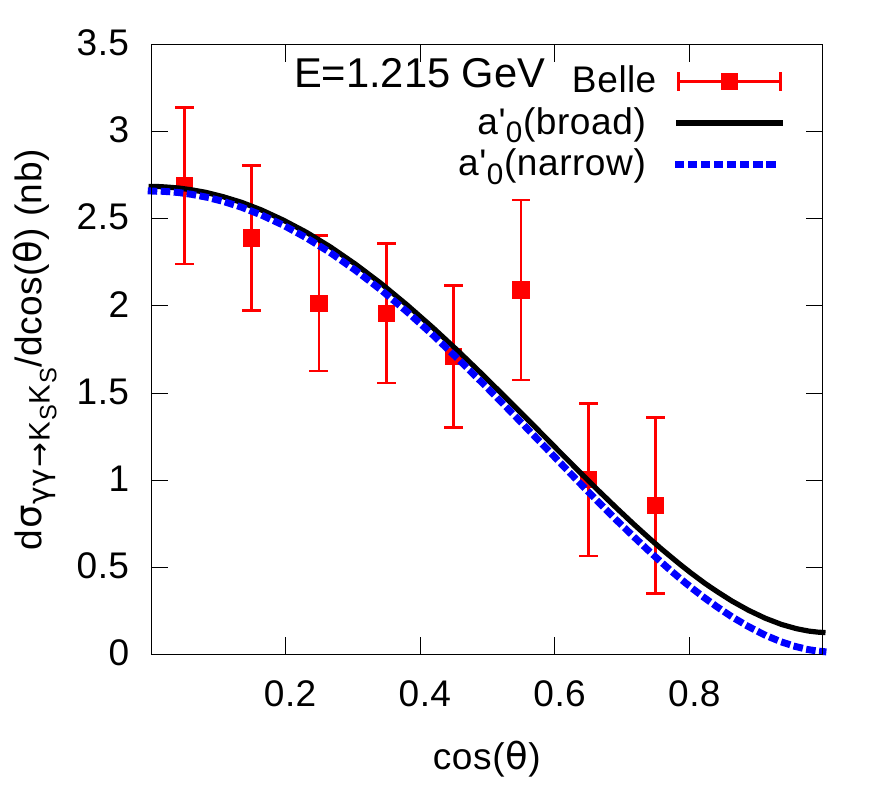}\\
\includegraphics[width=0.333\linewidth]{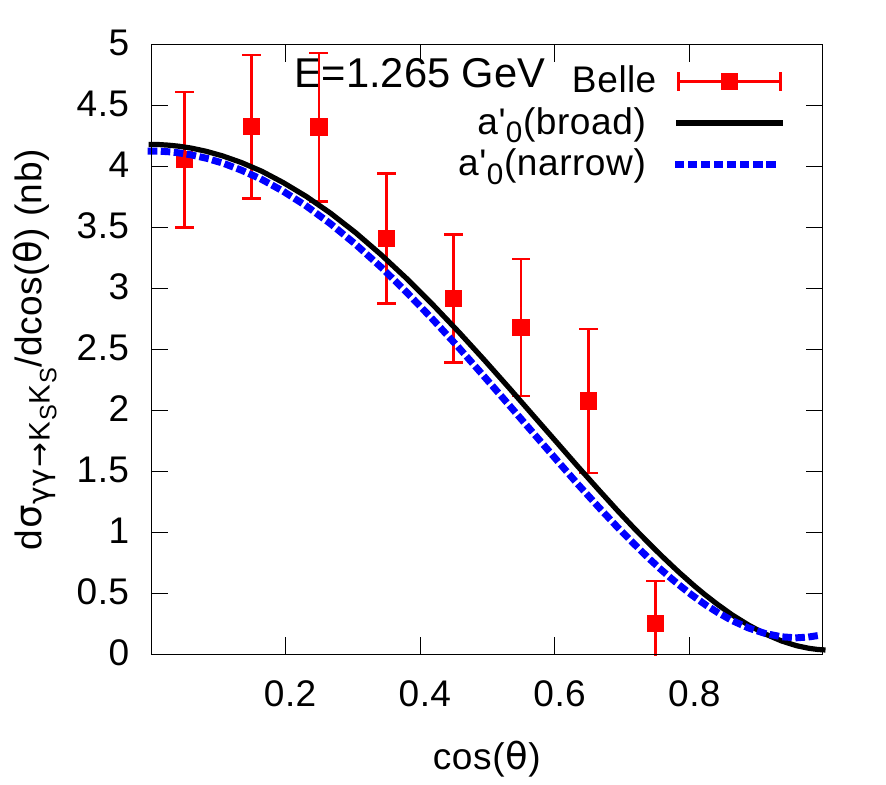}\includegraphics[width=0.333\linewidth]{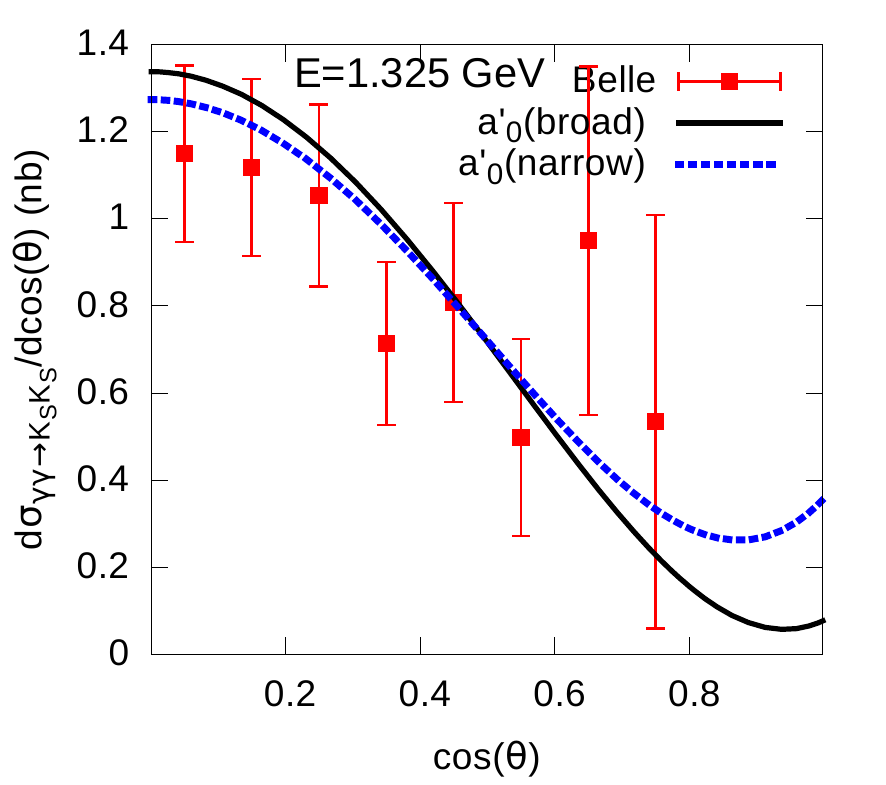}\includegraphics[width=0.333\linewidth]{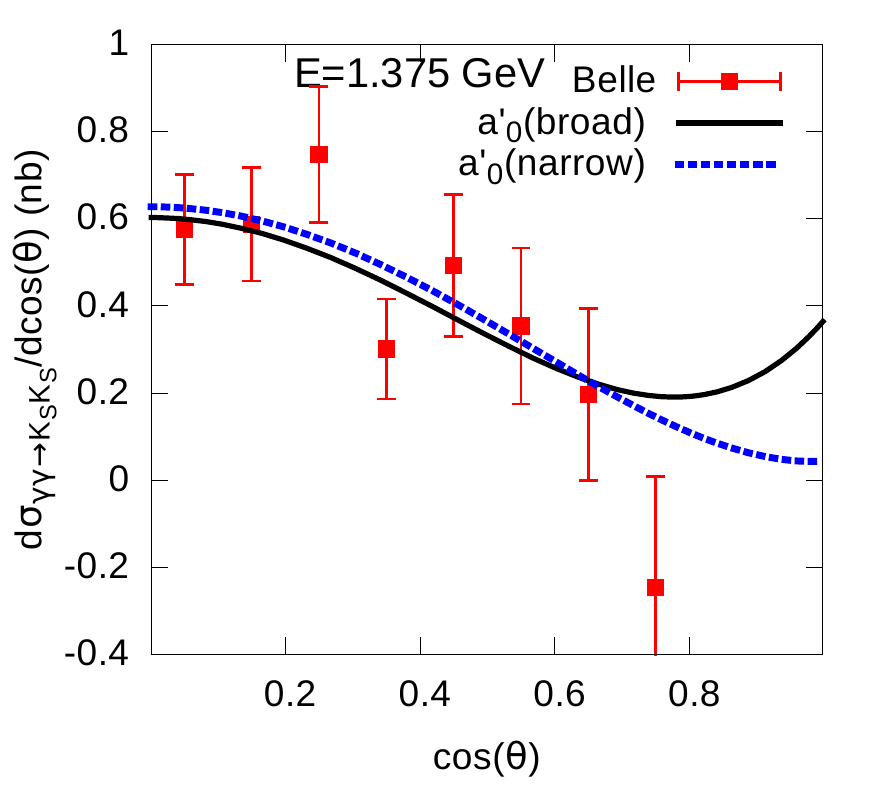}\\
\caption{\sl Experimental $\gamma\gamma\to K_SK_S$ differential
  cross-sections compared with the two fits results.}
\lblfig{diffcrossKsKs}
\end{figure}

The values of the remaining seven parameters included in the fit (subtraction
parameter $b_k$, tensor mesons to $2\gamma$ coupling constants, $a_2$ mass and
width) are collected in table~\ref{tab:tensorcoupl} below. The couplings
$D^T_{2\gamma}$ are found, as expected, to be smaller in magnitude than the
$C^T_{2\gamma}$ although this suppression is only by a factor of two in the
case of fit I. The values are in qualitative agreement with the PDG
expectations except, however, for the $a_2$ mass which is shifted\footnote{ A
  recent determination (from the complex pole) of the $a_2$ mass using data
  from the COMPASS experiment also finds a shift compared to the PDG value,
  however in the opposite direction\cite{Jackura:2017amb}.  } by approximately
10 MeV. This shift is easily seen to be caused by the presence of non-resonant
contributions to the $J=2$ amplitudes modelled here by the vector-meson
exchanges, $l^V_{2,\lambda\lambda'}$. The presence of this term is essential
for obtaining a correct description of the amplitude in the $\eta$ decay
region $s < (\meta-\mpi)^2$. In the 1 GeV energy region, one expects some
modifications induced by higher mass exchanges, but we see no reason that it
should be completely cancelled. We will consider a variation of this term as a
source of error below.

\begin{table}[b]
\centering
\begin{tabular}{c|c|c|c|c|c|c|c}
\hline\hline
\TT\BB & $C^{a_2}_{\gamma\gamma}$ & $D^{a_2}_{\gamma\gamma}$ &
     $C^{f_2}_{\gamma\gamma}$ & $D^{f_2}_{\gamma\gamma}$ & 
     $m_{a_2}$  & $\Gamma_{a_2}$ & $b_k$ \\[0.2cm] \hline
\TT
fit I & $0.105(2)$ &$\phantom{-}0.051(9)$ & $0.161(2)$ & $0.081(12)$ &
$1.328(1)$ & $0.097(2)$ & $-0.254(1)$ \\[0.2cm]
fit II  & $0.106(2)$ &$-0.033(6)$ & $0.171(2)$ & $0.007(9) $ &
$1.332(1)$ & $0.112(3)$ & $-0.167(5)$ \\
\hline\hline
\end{tabular}
\caption{\sl  Values of the coupling constants, of the subtraction constant
  $b_k$ and of the mass and width of the $a_2(1320)$ resonance resulting from
  the two fits.}\label{tab:tensorcoupl} 
\end{table}
\subsection{Properties of the $a_0$ resonances}
In this section we consider in more detail the properties of the two $a_0$
resonances which can be deduced from our analysis of the photon-photon
data. We will focus on the results from fit II since fit I was
argued not to be physically relevant. The formulas needed to define the
$T$-matrix elements on the unphysical Riemann sheets are given in
appendix~\sect{Tmatrix4sheets}. One may define coupling constants from the
residues of the resonance poles (e.g.~\cite{Morgan:1987gv} in the context of
photon-photon amplitudes), the couplings to the $\pi\eta$ and $K\Kbar$
channels are thus defined as
\be\lbl{couplingsresid}
\left.16\pi T_{11}^{(II)}(z)\right\vert_{pole}=
\frac{g^2_{a_0\pi\eta}}{z_{a_0}-z},\quad
\left.16\pi T_{12}^{(II)}(z)\right\vert_{pole}=
\frac{g_{a_0\pi\eta}g_{a_0K\Kbar}}{z_{a_0}-z}
\en
and similarly for the third Riemann sheet. The coupling to the $\gamma\gamma$
channel and the associated width are defined as
\be
\left. e^2l_{0++}^{(II)}(z)\right\vert_{pole}=
\frac{g_{a_0\gamma\gamma}g_{a_0\pi\eta}}{z_{a_0}-z},\quad
\Gamma_{a_0\to\gamma\gamma}= \frac{\vert g_{a_0\gamma\gamma}  \vert^2}
{16\pi m_{a_0}}\ .
\en
The following numerical results are found on sheet II for the position of the
$a_0(980)$ pole and the corresponding coupling
constants 
\be\lbl{a0980coupl}
\ba{ll}
\sqrt{s_{a_0}}=   & 1000.7(7)-i\,36.6(1.3)       \ (\MeV)\\[0.2cm]      
|g_{a_0\pi\eta}|=    & 2.17(2) \ (\hbox{GeV})\\[0.2cm]
|g_{a_0K\Kbar}| =    & 4.03(2)\ (\hbox{GeV})\\[0.2cm]
\Gamma_{a_0\gamma\gamma}=& 0.52(1)  \ (\hbox{keV})\ .
\ea\en
From the sheet III resonance, the position of the pole was given in
eq.~\rf{a01450pole} and the corresponding coupling constants have the
following values
\be\lbl{a01450coupl}
\ba{ll}
|g_{a'_0\pi\eta}|= &  3.15(4) \ (\hbox{GeV})\\[0.2cm]
|g_{a'_0K\Kbar} |= &  1.89(4) \ (\hbox{GeV})\\[0.2cm]
\Gamma_{a'_0\gamma\gamma}=& 1.05(5) \ (\hbox{keV})\ .
\ea\en

\begin{table}
\centering
\begin{tabular}{c|c|c|c|c|c||c }
\hline\hline
\TT\BB                    & $s_A$ & $L_i$ & $c'_m/c'_d$ & LC(a) &  LC(b)  &Total  \\
\hline
\TT $m_{a_0}$ (MeV)        &$[-0.2,1.9]$  &$5.7$  &$[4.2,10.8]$& $1.7$     &$[1.9,3.0]$&$[-0.2,12.9]$   \\[0.2cm]
$\Gamma_{a_0}/2$ (MeV)     &$[-2.3,3.7]$   &$3.8$ &$[7.1,11.0]$& $0.5$     &$[0.3,3.0]$&$[-2.3,12.6]$   \\[0.2cm]
$|g_{a_0\pi\eta}|$ (GeV)    &$[-0.15,0.21]$&$0.20$ &$[0.21,0.47]$ & $0.13$  &$[-0.03,0.19]$&$[-0.2,0.6]$   \\[0.2cm]
$|g_{a_0K\Kbar}|$ (GeV)     &$[-0.03,0.05]$&$-0.13$&$[0.01,0.28]$ & $0.05$  &$[-0.03,0.04]$& $[-0.2,0.3]$  \\[0.2cm]
\BB $\Gamma_{a_0\gamma\gamma}$ (keV)
                       &$[-0.05,0.07]$   &$0.07$  &$[0.06,0.16]$  & $0.14$ &$[-0.01,0.06]$&$[-0.1,0.2]$   \\    \hline
\TT $m_{a'_0}$    (MeV)    &$[2.4,30.2]$  &$104.2$ &$[22.2,62.3]$  &$-11.4$ &$[-12.9,53.3]$&$[-17,136]$   \\[0.2cm]
$\Gamma_{a'_0}/2$ (MeV)    &$[18.3,44.0]$ &$68.2$  &$[6.6,14.8]$   &$ 4.2$  &$[18.2,66.5]$&$[0,106]$     \\[0.2cm]
$|g_{a'_0\pi\eta}|$ (GeV)   &$[-0.68,0.01]$&$-0.28$ &$[-0.05,0.10]$ &$0.13$  &$[-0.89,0.10]$&$[-1.2,0.2]$    \\[0.2cm]
$|g_{a'_0K\Kbar}|$ (GeV)    &$[-0.26,0.06]$&$-0.20$ &$[-1.56,-1.42]$&$-0.05$ &$[-0.17,-0.07]$&$[-1.6,0.06]$   \ \\[0.2cm]
$\Gamma_{a'_0\gamma\gamma}$ (keV)
                      &$[0.14,0.22]$&$0.40$&$[-0.28,-0.25]$ & $-0.13$&$[0.21,0.17]$& $[-0.3,0.5]$\\ \hline\hline
\end{tabular}
\caption{\sl Errors generated by varying the fixed parameters of the $S$-wave
  amplitudes, columns 5 and 6 refer to the left-cut (see text).}\label{tab:listerrors}  
\end{table}
The errors quoted in the above formulas are those arising from the fitted
parameters (for which MINUIT provides a correlation matrix).  
In addition to those, one must consider errors  associated with
further parameters on which the $T$-matrix depends, which we have assumed to
be fixed up to now:
\begin{itemize}
\item[1)] Adler zero: the value of Adler zero was fixed to $s_A=\metad$ but its
  exact value is not known, we will vary it in the range $s_A=\metad\pm
  3\mpid$.  

\item[2)] Set of the $O(p^4)$ parameters $L_i$: in the BE14
  determination~\cite{Bijnens:2014lea}, their values are given with errors but
  there are strong correlations. In order to estimate an error from this
  source we used two different sets of $L_i's$ which correspond to two best
  fits made with different assumptions (second and third column in Table 3 of
  ref.~\cite{Bijnens:2014lea}). 

\item[3)] The ratio of the scalar couplings $c'_m/c'_d$ (see
  eqs.~\rf{K6},~\rf{K6b}) was varied between 1 and 3 to estimate an
  uncertainty from this source.

\item[4)] Left-hand cut: a) we added a contribution from a set of axial-vector
  mesons\footnote{The coupling constants of the $C$-odd axial-vectors are not
  very precisely known (see~\cite{Ko:1992zr}.). We made a simple estimate
  taking a mass $M_A=1.2$ GeV and an effective coupling
  $G_A=0.30(C_{\rho\pi}C_{\rho\eta}+C_{\omega\pi}C_{\omega\eta})$.} and b) we
  varied the products of the couplings of the vector mesons by $\pm20\%$. 
\end{itemize}
Table~\ref{tab:listerrors} gives the detailed list of the errors generated from
these variations on the properties of the two $a_0$ resonances.

\subsection{The $\eta\to \piz\gamma\gamma$ decay amplitude}
We consider now the modelled  amplitudes in the kinematical
region relevant for the decay $\eta\to \piz\gamma\gamma$. Including
the $J=0$ and $J=2$ partial-waves\footnote{Contributions from $J\ge4$
  partial-waves can be checked to be completely negligible.}, 
the distribution as a function of
the $\gamma\gamma$ invariant mass squared, $s$, reads
\be\lbl{etadistrib}
\ba{ll}
\dfrac{d\Gamma^{\eta\to\gamma\gamma \pi} }{ds}
= \dfrac{\alpha^2}{16\pi m_\eta^3} \sqrt{\lambda_{12}(s)}& \Big\{
\vert l_{0,++}(s)+\tilde{l}_{0,++}(s)\vert^2
+5\,\vert l^V_{2,++}(s) + l^{BW}_{2,++}(s)\vert^2\\[0.3cm]
\ & +5\,\vert l^V_{2,+-}(s) + l^{BW}_{2,+-}(s)\vert^2
\Big\}
\ea\en 
also accounting for the $J=0$ isospin violating amplitude
$\tilde{l}_{0,++}$. 
\begin{figure}
\centering
\includegraphics[width=0.6\linewidth]{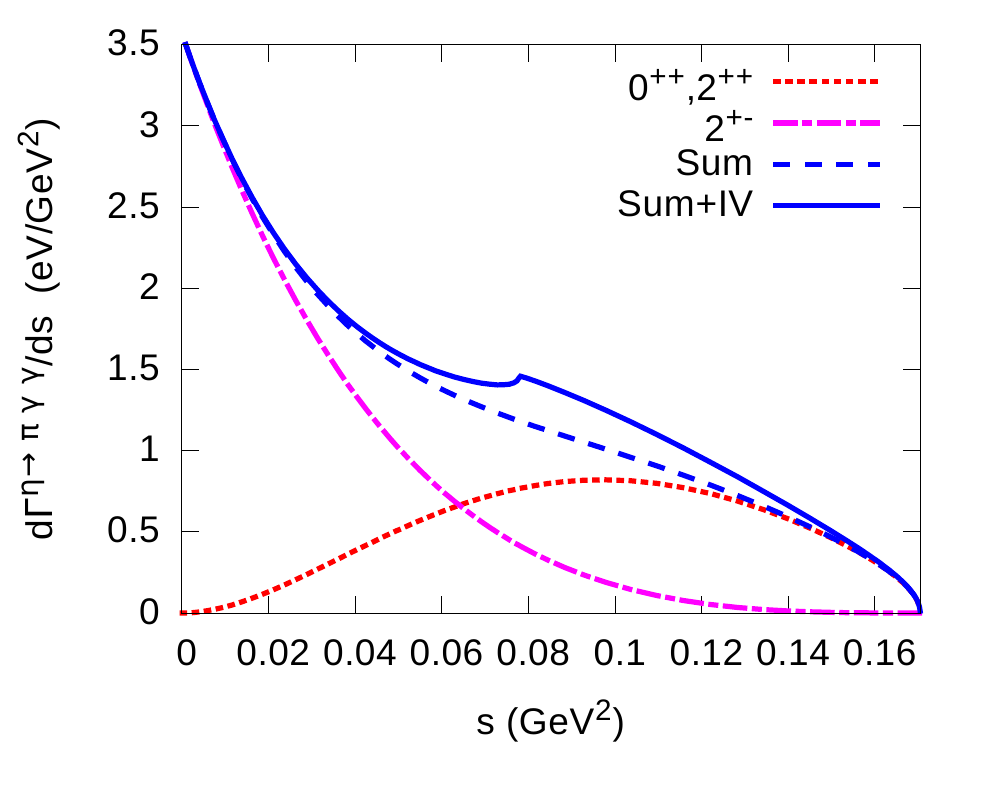}
\caption{\sl Contributions to the $\eta\to \pi^0\gamma\gamma$ energy
  distribution from the amplitudes fitted in the scattering
  region. Dotted line (red): $S$-wave (isospin conserving) and
  $2^{++}$ $D$-wave, dash-dotted line (magenta): $2^{+-}$ $D$-wave,
  dashed line: sum of the $S$ and $D$ waves, solid line: sum including
  the isospin violating amplitude $\tilde{l}_{0++}$. }
\lblfig{etadiffwidth}
\end{figure}
\begin{figure}
\centering
\includegraphics[width=0.6\linewidth]{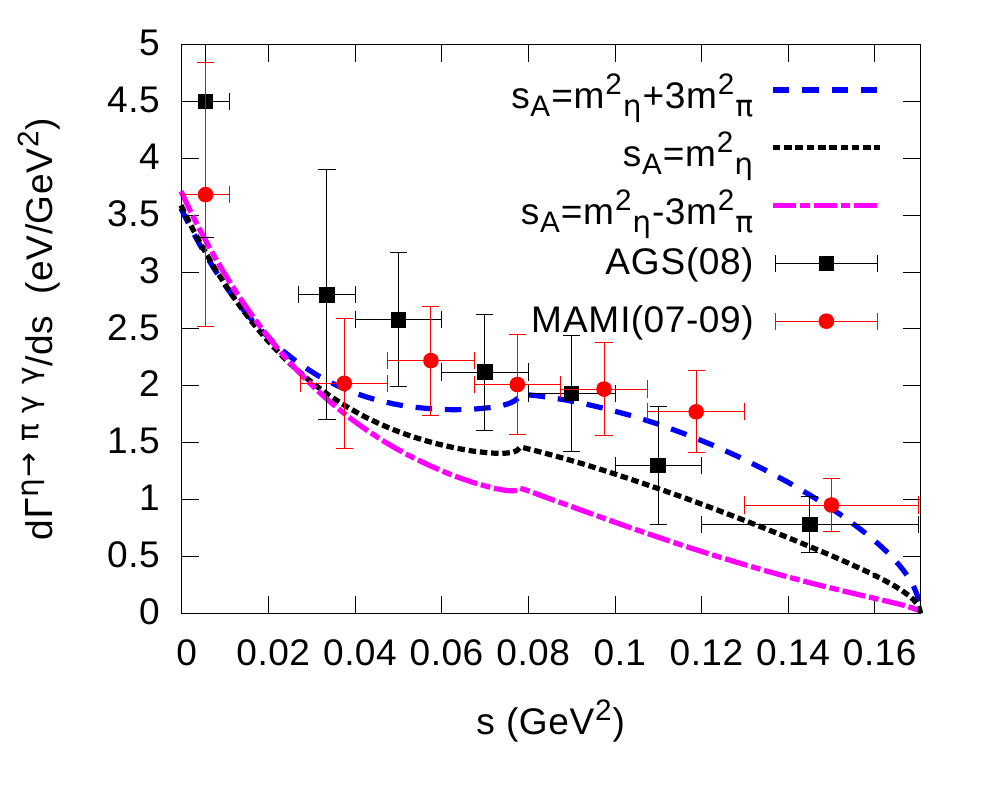}
\caption{\sl Experimental data on the $\eta\to \pi^0\gamma\gamma$ energy
  distribution~\cite{Prakhov:2008zz,Nefkens:2014zlt} compared with predictions
  from our amplitudes showing the influence of the position of the Adler zero.}
\lblfig{etadiffwidth1}
\end{figure}
Fig.\fig{etadiffwidth} illustrates the contributions of the various amplitudes
corresponding to central values of the parameters fitted in the scattering
region. The $2^{+-}$ partial-wave dominates over the other ones near $s=0$
because the $J^{++}$ amplitudes are suppressed by the soft-photon zero. The
relative role of the $S$-wave increases with the energy and starts to be
dominating above the $\pi^+\pi^-$ threshold. The figure also shows that the
isospin-violating $S$-wave generates a visible cusp at this threshold. The
central value of the decay width generated by our amplitudes is $\Gamma=0.237$
eV which is on the low side of the most recent experimental determinations
$\Gamma_{exp}=0.285\pm0.031\pm0.061$ eV (Crystal Ball at the
AGS~\cite{Prakhov:2008zz}), $\Gamma_{exp}=0.33\pm 0.03$ eV
(MAMI~\cite{Nefkens:2014zlt}). A smaller value was reported by the KLOE-2
collaboration~\cite{DiMicco:2005stk} but it has not been confirmed and the
data is currently being reanalysed. The amplitudes in the $\eta$ decay region
are very sensitive to the precise position of the Adler zero. This is
illustrated in fig.~\fig{etadiffwidth1} showing the effect of varying $s_A$ in
the range $[\metad-3\mpid,\metad+3\mpid]$ and comparing with the experimental
results. Given somewhat more precise data the value of $s_A$ could be included
in the fitting. The amplitudes in the decay region are also sensitive to the
vector meson coupling constants, the value of which dominate the energy
distribution near $s=0$. Accounting for these main errors we would predict
\be
\Gamma^{\eta\to \piz\gamma\gamma}= 0.237^{+0.060}_{-0.043}\ \hbox{eV}\ .
\en

\section{Conclusions}
In this work we have reconsidered the properties of the light
isovector scalar resonances as can be determined from photon-photon
scattering experimental results. For this purpose, we have implemented
a standard Muskhelishvili-Omn\`es integral representation for the $J=0$
amplitude in which the left-cut is modelled from light vector meson
exchanges. The underlying $T$-matrix satisfies unitarity with two
channels ($\pi\eta,K\Kbar$) and involves six phenomenological
parameters. In the case of the $J=2$ amplitudes the constraints from
unitarity are more difficult to implement, a cruder
description is used which consists in simply adding the cross-channel
vector-exchange and the direct channel tensor resonance amplitudes. In
order to constrain the free parameters as unambiguously as possible we
performed fits to both $\pi\eta$ and $K_SK_S$ data, for which
high-statistics data below 1.4 GeV are available at present, and we found two
different acceptable solutions to the minimisation. Both solutions are also
compatible with the available $K^+K^-$ data from the ARGUS
collaboration~\cite{Albrecht:1989re}. 

In one of our fits the $S$-wave amplitude displays a light and narrow $a'_0$
resonance exactly similar to the one found in the Belle analysis. While this
is mathematically allowed we have argued that the fit which displays a broad
$a'_0$ is likely to be more physical. 
Concerning the $a_0(980)$ resonance, we find that a rather conventional
picture i.e. a pole on the second sheet with a mass and width compatible with
the PDG and coupling to both the $\pi\eta$ and the $K\Kbar$ channels is
perfectly compatible with both the $\pi\eta$ and the $K_SK_S$ data. This is in
contrast with the Belle analysis which uses an elastic Breit-Wigner
description and also with the recent analysis of ref.~\cite{Danilkin:2017lyn}
in which the mass and width are found to be both significantly larger than the
PDG values. Data with a better energy resolution would be useful to resolve
these remaining ambiguities. The $\gamma\gamma\to \Kp\Km,K_SK_S$
cross-sections close to the $K\Kbar$ thresholds are also very sensitive to the
position of the $a_0(980)$.  Our results in this energy region are in
qualitative agreement with the chiral-unitary calculations from
ref.~\cite{Oller:1997yg} and with the estimates made in
ref.~\cite{Achasov:2012sc} but not with those from ref.~\cite{Dai:2014zta}.
Experimental data in this near-thresold region would obviously be very
constraining. 
Finally, it will be quite interesting to see how the pole
position determined in a lattice QCD simulation~\cite{Dudek:2016cru} evolves
when the value of $m_\pi$ is decreased.


\bigskip
\noindent{\large\bf Acknowledgements}\\
We would like to thank prof. S. Uehara for many clarifying explanations about
the Belle experiments. JXL thanks Prof. Lisheng Geng for useful
discussions. JXL is partly supported by the National Natural Science
Foundation of China under Grant Nos.11735003, 11975041, 11961141004 and by the
Fundamental Research Funds for the Central Universities. He also gratefully
acknowledges the financial support from the China Scholarship Council. This
work is supported in part by the European Union's Horizon2020 research and
innovation programme (HADRON-2020) under the Grant Agreement n$^\circ$ 824093.
\appendix
\appendix
\section{Scalar form factors}\label{sec:ffactors}
In this section we consider how the two isovector scalar form factors
get modified as compared to the results of
ref.~\cite{Albaladejo:2015aca} when using the set of $T$-matrix
parameters as determined here from the $\gamma\gamma$ data. The form
factors were defined as
\be
\ba{l}
B_0F_S^{\eta\pi}(s)=\braque{\eta\pip\vert\, \bar{u}d(0)\, \vert0}\\[0.2cm]
B_0F_S^{K\Kbar}(s)=\braque{\Kzb\Kp\vert\, \bar{u}d(0)\, \vert0}
\ea
\en
where $B_0$ is the chiral coupling proportional to the quark
condensate~\cite{Gasser:1984gg} 
\be
B_0=- \left.\braque{0\vert \bar{u}{u}\vert0}/
F^2_\pi\right\vert_{m_u=m_d=m_s=0}\ .
\en
Under the assumption of two-channel unitarity these form factors, we
recall, are simply related to the MO matrix,
\be
\bp
F_S^{\eta\pi}(s)\\[0.2cm]
F_S^{K\Kbar}(s)
\ep=
\bm{\Omega}_0(s)\times
\bp
F_S^{\eta\pi}(0)\\[0.2cm]
F_S^{K\Kbar}(0)
\ep\ .
\en
The values at $s=0$  are estimated from the chiral expansion at
order $p^4$.  In ref.~\cite{Albaladejo:2015aca} the set of $L_i$
values from ref.~\cite{Bijnens:2014lea} based on a $p^4$ fit was used
while here we preferred to use a set from a $p^6$ fit (BE14 set), which gives
\be
\ba{l}
F_S^{\eta\pi}(0)= 0.972 \\[0.1cm] 
F_S^{K\Kbar}(0)=  0.845\ .
\ea\en
Interesting quantities which are related to the scalar form factors
are the scalar radii which are proportional to their derivatives at
$s=0$
\be
\braque{r^2}_S^{P_1P_2}\equiv
6\dot{F}_S^{P_1P_2}(0)/{F}_S^{P_1P_2}(0)\ .
\en
At chiral order $p^4$ the scalar radii depend on a single low-energy
coupling, $L_5$. Using its value from the BE14 set,
$L^r_5=(1.01\pm0.06)\cdot10^{-3} $, gives
\be
\ba{l}
\left.\braque{r^2}_S^{\eta\pip}\right\vert_{p^4}= 
       0.067(7)\  \hbox{fm}^3\\[0.1cm] 
\left.\braque{r^2}_S^{K\Kbar}\right\vert_{p^4}=0.111(7)\ \hbox{fm}^3\ .
\ea\en

Fig.~\fig{FFactors} shows the absolute values of the form factors
using the BE14 chiral  couplings and the set of $T$-matrix parameters
fitted to the $\gamma\gamma$ data. The $a_0(980)$
resonance displays a smaller peak than in
ref.~\cite{Albaladejo:2015aca} and its position is slightly shifted,
reflecting the modified position of the pole. The $a'_0$ resonance
also differs: in the case of the ``narrow $a'_0$'' fit solution it shows up as
a dip in the $\eta\pi$ form factor while in the case of the ``broad $a'_0$''
solution its effect is hardly visible. Concerning the scalar radii, the
dispersive amplitudes give
\be
\ba{l}
\braque{r^2}_S^{\eta\pi}= 
\left(2.73^{+4.10}_{-2.60}\right)\cdot10^{-2}\  \hbox{fm}^2\\[0.2cm]
\braque{r^2}_S^{K\Kbar}=0.156^{+0.002}_{-0.029}\ \hbox{fm}^2\ .
\ea\en
These results correspond to fit solutions with a ``broad'' $a'_0$.
As compared with the the chiral $O(p^4)$ result, the central value is
somewhat too small in the case of $\eta\pi$ and too large in the case
of $K\bar{K}$.  

\begin{figure}[hbt]
\centering
\includegraphics[width=0.55\linewidth]{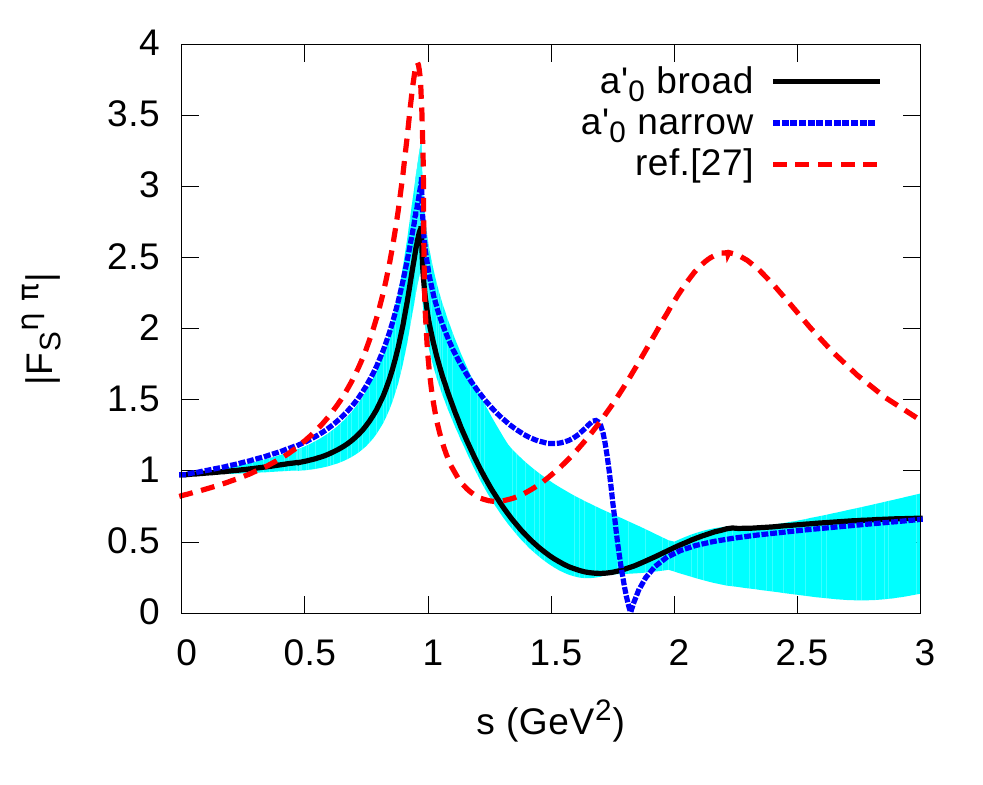}
\includegraphics[width=0.55\linewidth]{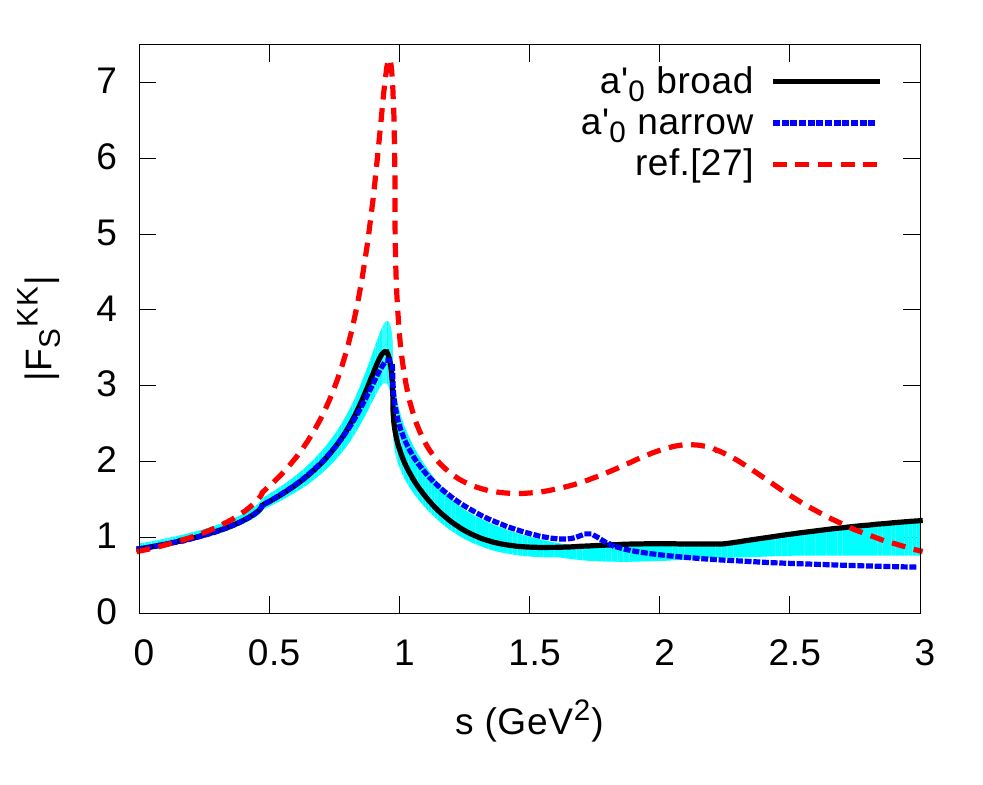}
\caption{\sl Absolute values of the scalar form factors: the results
  using the sets of $T$-matrix parameters from the two different fits
  to the $\gamma\gamma$ data are compared with a typical result from
  ref.~\cite{Albaladejo:2015aca}. The bands are obtained by varying
  the intrinsic parameters as in table~\ref{tab:listerrors}.}
\lblfig{FFactors}
\end{figure}
\section{Chiral expansion results for $\gamma\gamma\to P_1P_2$
  amplitudes}\label{sec:chiralrev} 
\subsection{Order $p^4$}
We collect below the expressions of the photon-photon amplitudes in the chiral
expansion at order $p^4$. The chiral order $p^2$ coincides with scalar QED and
gives rise to the Born amplitudes for $\pip\pim$ and $\Kp\Km$ which were given
in eq.~\rf{BornAmplit}. The leading contributions to the amplitudes which
involve two neutral mesons: $\pi^0\pi^0$, $\Kz\Kzb$ and $\piz\eta$ appear at
order $p^4$. They were computed in
refs.~\cite{Bijnens:1987dc,Donoghue:1988eea,Guerrero:1997rd,Ametller:1991dp}.
The basic one-loop function which occurs in these $p^4$ amplitudes can be
written as,
\be\lbl{G_P}
G_P(z)=-1 -\frac{m_P^2}{z}\left(
\log \dfrac{\sqrt{1-4m_P^2/z}+1}{\sqrt{1-4m_P^2/z}-1}
\right)^2\ ,
\en
for small $z$ values, $|z| << m_P^2$, it behaves as
\be
 G_P(z)=\dfrac{z}{12m_P^2} +O(z^2)\ .
\en
The expression for the $\gamma\gamma\to \pi^0\pi^0$ scalar amplitudes
at $O(p^4)$ read, 
\be\lbl{pi0pi0NLO}
\ba{l}
A_{NLO}^{\piz\piz}(s,t,u)=\dfrac{1}{4\pi^2\,s}\bigg[ 
\dfrac{s-\mpid}{\fpid}  G_\pi(s) +
\dfrac{s}{4\fpid}  G_K(s)\bigg] \\[0.2cm]
B_{NLO}^{\piz\piz}(s,t,u)=0\ .
\ea\en
These formulas illustrate general features: the $O(p^4)$ part of the $A$
amplitudes have a simple structure involving a sum of products of meson-meson
amplitudes at order $p^2$ with one-loop functions $G_P$ while the $B$
amplitudes vanish at this order. Next, the expression for $\gamma\gamma\to
\Kz\Kzb$ at $O(p^4)$ reads
\be\lbl{KzKzbNLO}
\ba{l}
A_{NLO}^{\Kz\Kzb}(s,t,u)=\dfrac{1}{4\pi^2\,s}\bigg[ 
\dfrac{s}{4\fpid}\, G_\pi(s) +
\dfrac{s}{4\fpid}\, G_K(s)\bigg]\ .\\
\ea\en
The NLO expression for the $\gamma\gamma\to \piz \eta$ amplitude, which is of
particular interest here, was worked out in ref.~\cite{Ametller:1991dp}. They
considered a chiral framework which also includes the $\eta'$ meson as a light
meson in addition to the pseudo-Goldstone octet. The relation between the
$\eta$ and $\eta'$ mesons and the octet and singlet chiral fields involves a
mixing angle $\theta$,
\be
\ba{l}
\eta=  c_\theta\, \phi^8 -s_\theta\, \phi^0\\
\eta'= s_\theta\, \phi^8 +c_\theta\, \phi^0
\ea\en 
denoting $c_\theta=\cos\theta$, $s_\theta=\sin\theta$. The expression for the
$\gamma\gamma\to \piz\eta$ amplitude as a function of $\theta$ reads,
\be\lbl{pi0etaNLO}
\ba{l}
A_{NLO}^{\piz\eta}(s,t,u)=\dfrac{1}{4\pi^2\,s}\bigg[ 
\dfrac{\sqrt3\, c_\theta}{36\fpid}\big( 9s-\mpid-8\mkd-3\metad
-\sqrt8 t_\theta(\mpid+2\mkd) \big)\, G_K(s)\\[0.3cm]
\quad+\dfrac{B_0(m_d-m_u)}{3\sqrt3(\metad-\mpid)\fpid}\big(
(c_\theta-\sqrt2 s_\theta)(4\mpid-3 s)
-t_\theta(4\sqrt2c_\theta+s_\theta)\mpid\big)\,G_\pi(s)\bigg]\\[0.5cm]
\ea\en
with $t_\theta=\tan\theta$. The isospin conserving part in eq.~\rf{pi0etaNLO}
agrees with ref.~\cite{Escribano:2018cwg}. It reproduces the result from
ref.~\cite{Ametller:1991dp} upon setting $s_\theta=-1/3$, $c_\theta=\sqrt8/3$
and neglecting the terms proportional to $t_\theta$ (which are not really
negligible).  In the standard ChPT one must set $c_\theta=1$, $s_\theta=0$ at
leading order. Doing so, the amplitude simplifies significantly and reads
\be\lbl{pi0etaNLOsimpl}
\ba{l}
A_{NLO}^{\piz\eta}(s,t,u)=\dfrac{1}{4\pi^2\,s}\bigg[ 
\dfrac{3s-4\mkd}{4\sqrt3\fpid}\,G_K(s)\\[0.4cm]
\quad +\dfrac{B_0(m_d-m_u)}{\metad-\mpid}\,
\dfrac{4\mpid-3s}{3\sqrt3\fpid}\big(G_\pi(s)-\frac{1}{2}G_K(s)\big)
\bigg]
\ea\en
where the Gell-Mann-Okubo mass relation was used and the kaon loop
contribution to the isospin violating part was included for completeness. 
The quantity $B_0(m_d-m_u)$ is given at leading chiral order  by
\be
B_0(m_d-m_u)=\left(m^2_\Kz- m^2_\Kp\right)_{QCD}\ . 
\en
Let us quote also the formula for $\gamma\gamma\to\eta\eta$
\be
\ba{l}
A^{\eta\eta}_{NLO}(s,t,u)=\dfrac{1}{4\pi^2 s} \bigg[
\dfrac{\mpid}{3\fpid}\,G_\pi(s) + \dfrac{9s-8\mkd}{12\fpid}\,G_K(s)
\bigg]\ .\\
\ea\en
Finally,  the order $p^4$ contributions to the $\gamma\gamma\to \pip\pim$
and $\Kp\Km$ amplitudes read 
\be
\ba{l}
A^{\pip\pim}_{NLO}(s,t,u)= \dfrac{8}{\fpid}(L_9^r +L_{10}^r)
+\dfrac{1}{4\pi^2\,s}\bigg[ 
 \dfrac{s}{2\fpid}\,G_\pi(s)
+\dfrac{s}{4\fpid}\,G_K(s)
\bigg]\\
\ea\en
and
\be\lbl{K+K-NLO}
\ba{l}
A^{\Kp\Km}_{NLO}(s,t,u)= \dfrac{8}{\fpid}(L_9^r +L_{10}^r)
+\dfrac{1}{4\pi^2\,s}\bigg[ 
 \dfrac{s}{4\fpid}\,G_\pi(s)
+\dfrac{s}{2\fpid}\,G_K(s)
\bigg]\ .\\
\ea\en
Combining eqs.~\rf{KzKzbNLO} and~\rf{K+K-NLO} we get the
NLO part of the $\gamma\gamma\to (K\Kbar)^{I=1},\,(K\Kbar)^{I=0}$
amplitudes
\be\lbl{KKbar1NLO}
\ba{l}
A_{NLO}^{(K\Kbar)^1}=-\dfrac{4\sqrt2}{\fpid}(L_9+L_{10})-\dfrac{\sqrt2\,}
{4\pi^2\,s}\bigg[ \dfrac{s}{8\fpid} G_K(s) \bigg]\\[0.3cm]
A_{NLO}^{(K\Kbar)^0}=-\dfrac{4\sqrt2}{\fpid}(L_9+L_{10})-\dfrac{\sqrt2\,}
{4\pi^2\,s}\bigg[\dfrac{s}{4\fpid} G_\pi(s) 
+\dfrac{3s}{8\fpid} G_K(s) \bigg]\ .
\ea\en

\subsection{Soft pion limit (isospin conserving part)}
Let us consider a limit when the pion in the $\gamma\gamma\to \piz \eta$
amplitude becomes ``soft'' i.e. $p_1=0$. This limit is unphysical but it
allows to obtain exact results which should hold for the physical amplitude
modulo $O(\mpid)$ corrections. In this limit, firstly, one has
\be
s=\metad,\quad t=u=0\quad(\hbox{soft pion limit})\ ,
\en
and using standard soft pion methods one shows that the amplitude
$\gamma\gamma\to \piz \eta$ vanishes exactly at this point.
When $p_1=0$ the two tensors $T^{\mu\nu}_1$, 
$T^{\mu\nu}_2$ become degenerate 
\be
T^{\mu\nu}_2=2\metad \,T^{\mu\nu}_1
\en
such that the soft limit amplitude, in tensorial form, reads
\be\lbl{softcomb}
\lim_{p_1\to 0} W^{\mu\nu}(q_1,q_2; p_1,p_2)
= (A(\metad,0)+ 2\metad B(\metad,0))\,T_1^{\mu\nu} =0\ .
\en
The combination which appears in eq.~\rf{softcomb} is the helicity amplitude
$L_{++}$ in the soft pion limit.  This implies that the physical
helicity amplitude $L_{++}$ with $t=u$ should also have an Adler
zero as a function of $s$,
\be
L_{++}(s_A,t=u)=0
\en
with $s_A=\metad+O(\mpid)$. Let us consider the implication of this result for
the $j=0$ partial wave. The expansion of the helicity amplitude with $\cos\theta=0$
(which corresponds to $t-u=0$) reads
\be
L_{++}(s,\cos\theta=0)=\sum_{j\ even} (2j+1)\, (-1)^{j/2}\frac{(j-1)!! }{j!!}\,
l_{j,++}(s)\ .
\en 
When $s=\metad$, the amplitudes with $j\ge 2$ are suppressed by the
angular momentum barrier factor: $l_{2,++}(\metad)=O(\mpid)$,
$l_{4,++}(\metad)=O(\mpiq), \cdots$. The soft pion condition therefore
implies that the $j=0$ partial-wave should satisfy $l_{0,++}(\metad)=O(\mpid)$
such that an Adler zero should be present in this partial-wave amplitude,
\be
l_{0,++}(s_A)=0\ .
\en

\begin{figure}
\centering
\includegraphics[width=0.49\linewidth]{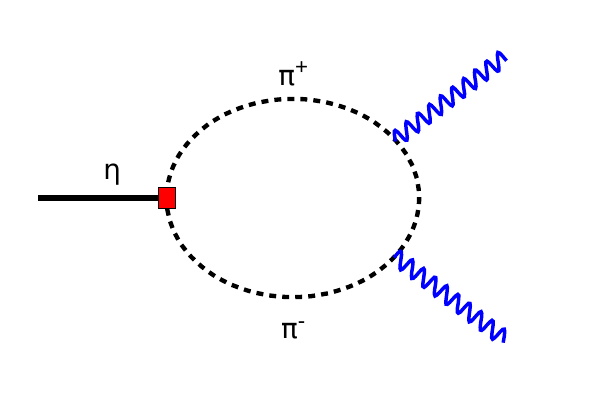}\includegraphics[width=0.49\linewidth]{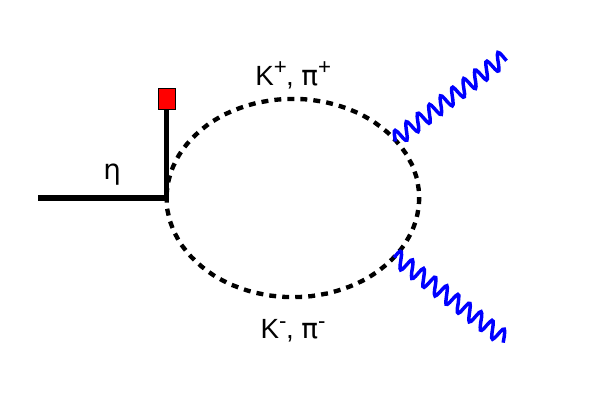}
\caption{\sl Representative chiral one-loop diagrams contributing to the
  matrix element of the operator $p_0=i(\bar{u}\gamma^5{u}+\bar{d}\gamma^5{d})
  $ needed in the soft-pion relation~\rf{softpirel}. Insertions of $p_0$ are
  represented by a red square.}
\lblfig{diagssoft}
\end{figure}
\subsection{Soft pion limit (isospin violating part)}\label{sec:l0tildeapp}
In the case of the isospin violating piece of the
$\gamma\gamma\to \piz \eta$ amplitude, the soft pion limit does
\emph{not} lead to an Adler zero. Let us define this
amplitude in terms of the following matrix element,
\be
\tilde{L}_{++}\equiv
\braque{\piz(p_1)\eta(p_2)\vert \frac{1}{2}(m_d-m_u)(\bar{u}u-\bar{d}{d})
\vert\gamma(q_1,+)\gamma(q_2,+)}\ ,
\en 
which involves the isospin violating part of the
QCD Lagrangian. Using the usual soft pion techniques, the soft pion
limit of this amplitude is expressed in terms of the commutator
\be
\lim_{p_1\to 0}\tilde{L}_{++}=-\dfrac{i}{2\fpi}
\braque{\eta(p_2)\vert (m_d-m_u)[Q_5^3, (\bar{u}u-\bar{d}{d})]
\vert\gamma(q_1,+)\gamma(q_2,+)}\ .
\en
where $Q_5^3$ is the axial charge
\be
Q_5^3=\frac{1}{2}\int d^3x 
(\bar{u}\gamma^0\gamma^5{u}-\bar{d}\gamma^0\gamma^5{d})(0,x)\ .
\en
The commutator is easily worked out and gives
\be\lbl{softpirel}
\lim_{p_1\to 0}\tilde{L}_{++}=\dfrac{(m_d-m_u)}{2\fpi}
\braque{\eta(p_2)\vert 
i(\bar{u}\gamma^5{u}+\bar{d}\gamma^5{d})
\vert\gamma(q_1,+)\gamma(q_2,+)}\ .
\en
The matrix element which appears on the right-hand side is a
non-perturbative quantity. We can estimate it using the chiral
expansion at NLO. The pseudoscalar operator
$p_0=i(\bar{u}\gamma^5{u}+\bar{d}\gamma^5{d}) $ can couple either to
a single $\eta$ meson or to $\eta\pi^+\pi^-$. Representative one-loop diagrams
are shown in fig.~\fig{diagssoft}. Computing them gives
\be\lbl{softmatrix}
\braque{\eta(p_2)\vert p_0(0)\vert
\gamma(q_1,+)\gamma(q_2,+)}_{NLO}=
\dfrac{-B_0}{4\pi^2\,\sqrt3\fpi }\, \big(1-\dfrac{\mpid}{3\fpid}\big)
\,\big(G_\pi(\metad)-\frac{1}{2} G_K(\metad)\big)\ . 
\en  
As one might expect, using this result in the soft pion
relation~\rf{softpirel} reproduces the isospin violating amplitude at
chiral order $p^4$ when $s=\metad$ (see eq.~\rf{pi0etaNLOsimpl}), up
to $O(\mpiq/m^4_\eta)$ terms. In practice, one can relate
$(m_d-m_u)B_0$ to the QCD kaon mass difference
$(m^2_\Kz-m^2_\Kz)_{QCD}$ and then use eq.~\rf{epsilonL} in order to
express $\tilde{l}_{0++}(\metad)$ in terms of $\epsilon_L$. The chiral
calculation~\rf{softmatrix} of the matrix element should be reliable
at the $20-30\%$ level since there are no potentially large $p^6$
contributions from light vector meson exchanges in this case.

\begin{figure}
\centering
\includegraphics[width=0.70\linewidth]{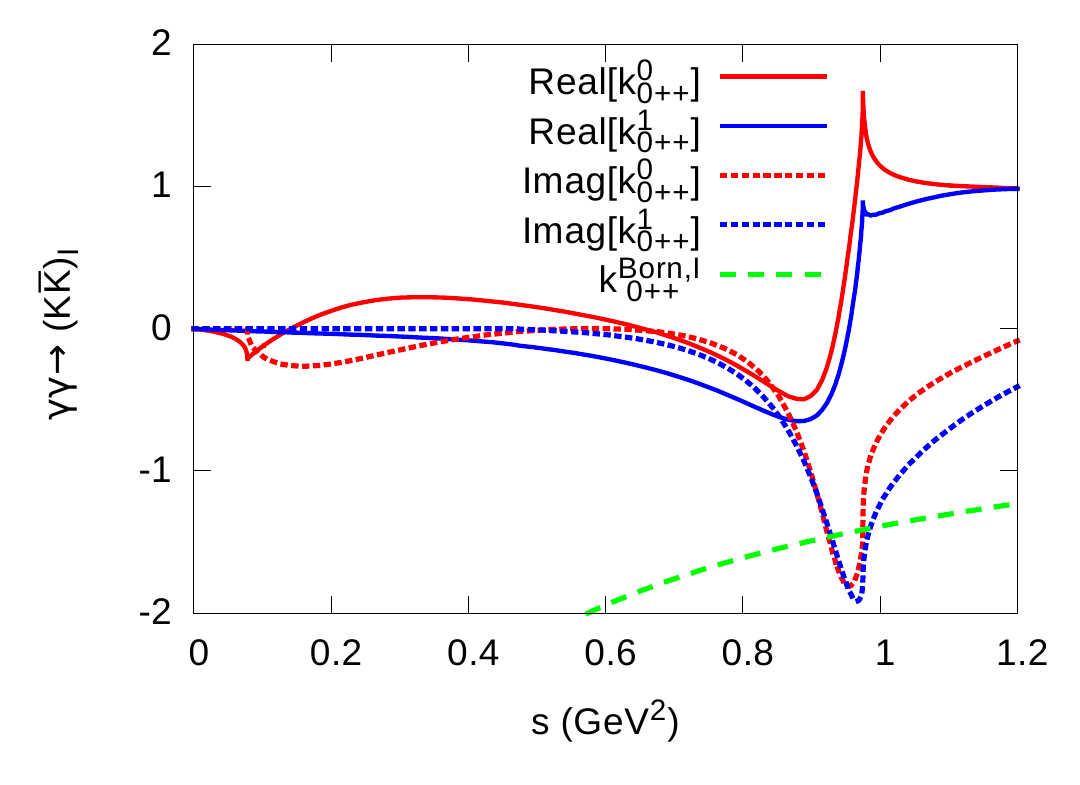}
\caption{\sl Dispersive $\gamma\gamma\to (K\Kbar)_I$ $S$-waves with $I=0,1$
without the Born amplitude which is shown separately.}
\lblfig{KKamplit}
\end{figure}

\section{Dispersive $\gamma\gamma\to \pi\pi,\ (K\Kbar)_{I=0}$
  $S$-waves}\label{sec:amplitI=0} 
Input for the $\gamma\gamma\to (K\Kbar)_{I=0}$ $S$-wave is needed in order to
reconstruct the amplitudes for the physical $K^+K^-$ and $K_SK_S$ states. We
briefly recall the dispersive coupled-channel representation for the $I=0$
$S$-wave taken from ref.~\cite{GarciaMartin:2010cw}. It has the following form
\be\lbl{omnesI=0}
\bp
h_{0++}(s)\\[0.3cm]
k^0_{0++}(s)\ep=
\bp
h^{Born}_{0++}(s)\\[0.3cm]
k^{0,Born}_{0++}(s)\ep+ \bm{\Omega}^{(0)}(s)\bp
b^0\,s +b^{'0}\,s^2 +L^{(0)}_1(s) + R^{(0)}_1(s)\\[0.3cm]
b^0_K\,s +b^{'0}_K\,s^2+ L^{(0)}_2(s) + R^{(0)}_2(s)
\ep
\en
which is analogous to eqs.~\rf{omnesSwaves} for the $I=1$ amplitudes but more
subtraction parameters had to be introduced. In eq.~\rf{omnesI=0}
$\bm{\Omega}^{(0)}$ is the $I=0$ Omn\`es matrix and the functions
$L^{(0)}_i(s)$ ($R^{(0)}_i(s)$) are dispersive integrals on the left
(right)-hand cuts which are analogous to the corresponding integrals in
eqs.~\rf{omnesSwaves}.  The parameter $b^0_K$ was simply fixed by matching to
the $O(p^4)$ amplitude at $s=0$ (see eq.~\rf{KKbar1NLO}).  The three
parameters $b^0$, $b^{'0}$, $b^{'0}_K $ where then determined from fits to
$\gamma\gamma\to \pip\pim, \piz\piz$ data. Having done so the $K\Kbar$
amplitude $k^0_{0++}$ was generated, which we have used in the present work
together with the $I=1$ amplitude.  Fig.~\fig{KKamplit} shows the dispersive
results for these two $K\Kbar$ amplitudes. At low energies the amplitude
$k^0_{0++}$ is significantly larger in magnitude than $k^1_{0++}$ because it
gets isoscalar $\pi\pi$ rescattering contributions (which contain the broad
$\sigma$ resonance). In the low-energy region, furthermore, the Born amplitude
is much larger than the rescattering contributions, in accordance with the
chiral counting. In the 1 GeV region the $I=0,1$ rescattering amplitudes are
rather similar which leads to a strong suppression of the $\Kz\Kzb$
cross-section as compared to the $\Kp\Km$ one close to the $K\Kbar$ threshold.

\section{The $T$-matrix on the four Riemann sheets}\label{sec:Tmatrix4sheets}
The second sheet extension of a matrix elements $T_{ij}$ is defined such as
to continue $T_{ij}(z)$ analytically  across the cut, below the first
inelastic threshold,
\be
T^{(II)}_{ij}(s-i\epsilon)= T_{ij}(s+i\epsilon),\quad 
(\meta+\mpi)^2 \le s \le 4\mkd\ .
\en
Using the elastic unitarity relation it easy to find an explicit expression
for $T^{(II)}_{ij}$,
\be\lbl{TIIexpl}
\bm{T}^{(II)}(z)=\left(  1-2\bm{T}(z)\begin{pmatrix}
\tilde{\sigma}_{\pi\eta}(z) & 0\\
0                  & 0\\ \end{pmatrix}\right)^{-1} 
\bm{T}(z)
\en
with 
\be
\tilde{\sigma}_{\pi\eta}(z)=\frac{\sqrt{ (z-m_-^2)(m_+^2-z) }}{z}\ . 
\en
The second sheet
extensions of the $\gamma\gamma$ amplitudes $l_{0++}$ and $k_{0++}$ are
expressed in a similar way,
\be
\begin{pmatrix}
l_{0++}^{(II)}(z)\\[0.2cm]
k_{0++}^{(II)}(z)\end{pmatrix}
=\left(  1-2\bm{T}(z)\begin{pmatrix}
\tilde{\sigma}_{\pi\eta}(z) & 0\\[0.2cm]
0                  & 0\\ \end{pmatrix}\right)^{-1}  
\begin{pmatrix}
l_{0++}(z)\\[0.2cm]
k_{0++}(z)\end{pmatrix}\ .
\en
The third sheet extension of the $T$-matrix elements are defined
such as to obey continuity equations across the unitarity cut above $4\mkd$
\be
T_{ij}^{(III)}(s-i\epsilon)=T_{ij}(s+i\epsilon),\quad 4\mkd \le s < \infty\ .
\en
They are  easily expressed in matrix form
\be\lbl{TIIIexpl}
\bm{T}^{(III)}(z)=\left(  1
-2\bm{T}(z)\begin{pmatrix}
\tilde{\sigma}_{\pi\eta}(z) & 0\\
0                  & \tilde{\sigma}_{KK}(z)\\ \end{pmatrix}
\right)^{-1}\,\bm{T}(z)
\en
where $\tilde{\sigma}_{KK}(z)=\sqrt{4\mkd/z-1}$. Finally, the fourth sheet
extensions of the $T$-matrix elements are given by
\be\lbl{TIVexpl}
\bm{T}^{(IV)}(z)=\left(  1
-2\bm{T}(z)\begin{pmatrix}
0                  & 0\\
0                  & \tilde{\sigma}_{KK}(z)\\ \end{pmatrix}
\right)^{-1}\,\bm{T}(z)
\en
they satisfy continuity equations with $T^{(II)}_{ij}(z)$ when $s > 4\mkd$ and
with  $T^{(III)}_{ij}(z)$ when $(\meta+\mpi)^2< s < 4\mkd$.

In our model, the right-cut structure of  $T$-matrix is generated by the
pair of loop functions $\bar{J}_{\pi\eta}(s)$,  $\bar{J}_{KK}(s)$
which can be written as 
\be\lbl{Jbardef}
\bar{J}_{12}(z)\equiv \bar{J}^I_{12}(z)
=\frac{z}{16\pi^2}\int_{(m_1+m_2)^2}^\infty ds'\,
\frac{\sqrt{\lambda_{12}(s')}}{(s')^2(s'-z)}\ . 
\en
This expression defines $\bar{J}_{12}$ on the first Riemann sheet and
shows that it is analytic  except for a cut on
$[(m_1+m_2)^2,\infty]$. The second sheet extension of $\bar{J}_{12}$ is given by
\be
\bar{J}^{II}_{12}(z)= \bar{J}_{12}(z)
+2\,\frac{\tilde{\sigma}_{12}(z)}{16\pi }\ .
\en
which satisfies the continuity equation $\bar{J}^{II}_{12}(s-i\epsilon)=
\bar{J}_{12}(s+i\epsilon)$ when $s \ge (m_1+m_2)^2$.  The $T$ matrix on the
four Riemann sheets can then also be defined as follows in terms of the pair
of loop functions
\be
\ba{ll}
T^{I}(z)   : & (\bar{J}^{I}_{\pi\eta}(z), \bar{J}^{I}_{KK}(z))\\[0.2cm]
T^{II}(z)  : & (\bar{J}^{II}_{\pi\eta}(z), \bar{J}^{I}_{KK}(z))\\[0.2cm]
T^{III}(z)  : & (\bar{J}^{II}_{\pi\eta}(z), \bar{J}^{II}_{KK}(z))\\[0.2cm]
T^{IV}(z)  : & (\bar{J}^{I}_{\pi\eta}(z), \bar{J}^{II}_{KK}(z))\ .\\
\ea\en
Using the $K$-matrix type representation~\rf{chiralK} of the $T$-matrix one
easily verifies that these definitions satisfy the relevant
equations~\rf{TIIexpl},~\rf{TIIIexpl},~\rf{TIVexpl} written above.


\end{document}